\documentclass{aa}  

\usepackage{graphicx}

\usepackage{txfonts}
\usepackage{amsmath}
\usepackage{amssymb}

\usepackage[colorlinks=true,linkcolor=blue,citecolor=blue,urlcolor=black]{hyperref}
\begin{document} 

   \title{Automated identification of dipole mixed modes in red-giant stars}

   \author{Francisca Espinoza-Rojas\inst{1,2}
           \and
           Michaël Bazot\inst{1,2}
           \and
           Saskia Hekker\inst{1,2}  
           \and
           Sébastien Bourguignon\inst{3} 
          }

   \institute{Heidelberger Institut für Theoretische Studien, Schloss-Wolfsbrunnenweg 35, D-69118 Heidelberg, Germany \\
              \email{francisca.espinoza@h-its.org}, 
              \and
              Center for Astronomy (ZAH/LSW), Heidelberg University, Königstuhl 12, D-69117 Heidelberg, Germany
              \and
             Laboratoire des Sciences du Numérique de Nantes, École Centrale de Nantes, 1 Rue de la Nöe, 44321 Nantes, France\\
             }

   \date{Received mm dd, yyyy; accepted mm dd, yyyy}

   \abstract
   {Dipole mixed modes probe the internal structure of red giants by carrying information from both the convective envelope and the radiative core. Consequently, identifying these modes allows for the determination of evolutionary states and physical conditions of the stellar interior. This identification typically relies on comparing observed frequencies with theoretical ones computed using the mixed-mode asymptotic relation. However, the process is challenging due to parameter-space degeneracies, stellar rotation, as well as missing modes and spurious peaks in the power density spectrum.} 
   {We aim to develop an automated and robust method for identifying dipole mixed modes in red giants, suitable for large-scale asteroseismic analyses.} 
   {We introduced a metric that measures the overlap between observed and asymptotic dipole frequencies represented as spike trains convolved with Gaussian kernels. After optimising this metric with a genetic algorithm, we used the best-fit parameters to perform a global assignment of the oscillation modes. We validated the method using synthetic frequencies of red-giant-branch and core-helium-burning stars perturbed by stellar rotation. Furthermore, we tested the method on four benchmark stars observed by \emph{Kepler}: KIC~4448777, KIC~5000307, KIC~9145955, and KIC~12008916.} 
   {Validation using perturbed synthetic frequencies demonstrates that our strategy recovers over 95\% of the mixed modes across different evolutionary states. This performance remains robust against common observational challenges such as spurious peaks and missing modes. For the benchmark stars, our recovered dipole mode identification is consistent with previously published results. Finally, our estimates for the parameters of the mixed-mode equation typically agree within $1\%$ for the period spacing ($\Delta\Pi_{1}$) and within $10\%$ for rotational splittings ($\delta\nu_{\rm rot}$) and coupling strength ($q$).}
   {Our analysis demonstrates that the proposed framework effectively addresses the challenges linked to the complex dipole-mode spectrum. Hence, it allows for the automated identification of dipole mixed modes, enabling homogeneous, large-scale asteroseismic analyses of evolved stars.} 
   
   
   \maketitle

\section{Introduction}\label{sec:intro}
   Solar-like oscillations in red-giant stars offer a direct window into stellar interiors. Excited stochastically by turbulent near-surface convection \citep{Garcia19}, these oscillations probe both the convective envelope and the radiative core. Identifying the oscillation modes in the power density spectrum (PDS) is therefore a critical step as it enables the extraction of seismic diagnostics that constrain stellar structure and evolution \citep[e.g.][]{Mosser12,Corsaro15b,GSOM18,Gehan18,Kallinger19,Corsaro20,Nielsen21,LiGang24,Nielsen25}. 

   Among the most informative seismic diagnostics in red giants are mixed modes ($\ell\geq1$), which arise from the coupling between pressure modes ($p$ modes) propagating in the envelope and gravity modes ($g$ modes) propagating in the core. Dipole ($\ell=1$) mixed modes are particularly valuable because they show the overall highest sensitivity to the core. Their identification has led to key advances in red-giant asteroseismology such as the distinction between hydrogen-shell-burning red-giant-branch (RGB) stars and core-helium-burning (CHeB) stars \citep[e.g.][]{Bedding11, Mosser12c,Kallinger12,Vrard16}. Furthermore, estimates of the rotational splittings of these modes have made it possible to infer the internal stellar rotation rates \citep[e.g.][]{Beck12,Deheuvels12,Deheuvels14,Deheuvels15,DiMauro16,Triana17,Tayar19,LiGang24}. These rotation rates in turn provide empirical constraints on angular momentum transport mechanisms across stellar evolution \citep[e.g.][]{Eggenberger12,Marques13,Cantiello14,Corsaro17,Mosser18,Moyano22,Eggenberger22,Bordadagua25}. 
 
   The analysis of red-giant oscillation spectra is, however, challenging due to their intrinsic complexity. Pure acoustic modes ($\ell = 0$) follow a regular $p$-mode pattern characterised by the large frequency separation ($\Delta\nu$) between consecutive radial orders \citep[e.g.][]{Mosser11UP}. In contrast, dipole modes are unevenly spaced in frequency as a consequence of their coupled $p$-$g$ character. Stochastic excitation and large variations in mode inertia lead to significant differences in amplitude between modes, causing some $g$-dominated modes to fall below the detection threshold. In addition, spurious peaks may contaminate the spectra and further complicate their interpretation. From a theoretical perspective, the mixed-mode frequency pattern is inherently degenerate with respect to variations in the $g$-mode period spacing $(\Delta\Pi_{1})$, phase offset $(\epsilon_{g})$, and the coupling strength $(q)$, resulting in a multi-modal parameter space \citep[e.g.][]{Buysschaert16,Kuszlewicz23}. Together, these observational and theoretical effects hinder the automated identification of dipole mixed modes, particularly in cases of incomplete or ambiguous mode detection.
   
   The availability of long, high-precision photometric time series from space missions such as CoRoT \citep{CoRoT}, \emph{Kepler} \citep{Borucki10}, K2 \citep{Howell14}, and the Transiting Exoplanet Survey Satellite \citep[TESS;][]{Ricker14} has enabled the detection of oscillations in tens of thousands of red giants \citep[e.g.][]{Yu18, Hon19, APOKASC3}, motivating the development of automated analysis techniques. While global asteroseismic parameters (i.e. the frequency of maximum oscillation power, $\nu_{\rm max}$, and $\Delta\nu$) are available for around 16,000 red giants \citep[e.g.][]{Yu18,APOKASC3}, the identification of their individual mixed modes remains a challenge.

   Different methods have been developed to characterise and identify dipole mixed modes in red-giant stars. \citet{Mosser09} introduced envelope autocorrelation techniques to detect regularities in oscillation spectra. This method was later extended to identify dipole modes and estimate rotational splittings \citep{Mosser12b}. Based on the asymptotic description of mixed modes, \citet{Mosser15} proposed the stretched-period formalism, which maps mixed-mode frequencies into a quasi-regular pattern in period. This approach has been widely applied to determine $\Delta\Pi_{1}$, the coupling factor $q$, and rotational splittings in large samples of RGB stars \citep[e.g.][]{Vrard16, Gehan18}. Moreover, it has been extended to measure core rotation in CHeB stars \citep[e.g.][]{Mosser24} and internal magnetic fields \citep[e.g.][]{LiGang22}. Other approaches infer these parameters by performing Bayesian optimisation directly on mode frequencies \citep[e.g.][]{Kuszlewicz23, LiGang24}. 
   
   Additionally, automated peak-bagging pipelines have been developed to extract and identify oscillation modes across large datasets. For example, the FAMED pipeline \citep{Corsaro20} combines multi-modal nested sampling with Bayesian model comparison to detect and label $\ell=1$ modes in a data-driven way. In contrast, {\tt PBjam} \citep{Nielsen21}, which was recently extended to $\ell=1$ modes \citep{Nielsen25}, relies on explicit mode-coupling models and non-parametric priors constructed mainly from previous measurements of thousands of stars to inform the Bayesian inference of dipole-mode frequencies.

   While these methods have enabled the detailed seismic characterisation of red-giant stars, observational limitations --- combined with the multi-modal nature of the mixed-mode parameter space --- challenge their applicability across the red-giant population. Specifically, purely data-driven approaches struggle to resolve mode patterns in spectra with dense dipole modes, spurious peaks, or low signal-to-noise ratios. Furthermore, Bayesian frameworks that rely heavily on empirical priors may not properly characterise stars whose properties fall outside the reference sample. This motivates alternative methods that assess the agreement between theoretical predictions and observations on a more global level. 
   
   In this paper, we introduce a new, unsupervised, optimisation-based scheme for the identification of $\ell=1$ mixed modes in red giants. The method takes a list of dipole-mode candidate frequencies provided by any peak-bagging code as input. It is based on a metric that quantifies the overlap between convolved spike trains \citep[e.g.][]{Kass14} of these observed frequencies and the theoretical predictions. This metric was optimised using a genetic algorithm to efficiently explore the high-dimensional and degenerate parameter space of mixed modes. By evaluating the agreement between observations and theoretical predictions on a global level, the method is robust against missing modes and spurious peaks, and it reliably fits the asymptotic relation. Therefore, it is suited for automated application to large stellar samples.

   The paper is organised as follows. In Sect.~\ref{sec:mixed_modes_formalism} we present the asymptotic formulation adopted for dipole mixed modes. In Sect.~\ref{sec:score_description} we describe the overlap metric based on convolved spike trains. In Sect.~\ref{sec:optimisation_modeID} we explain how the metric was optimised and how the identification of mixed modes was performed. In Sect.~\ref{sec:validation} we describe how we validated our method by applying it to a comprehensive grid of synthetic spectra that mimic observations of stars in different evolutionary states. We show the application to four stars observed by \emph{Kepler} in Sect.~\ref{sec:real_stars}. Finally, we discuss the advantages and limitations of the approach in Sect.~\ref{sec:discussion}.
      
\section{Mixed-mode formalism} \label{sec:mixed_modes_formalism}
   Non-radial oscillation modes in red-giant stars exhibit a mixed character due to the coupling between $p$ modes propagating in the convective envelope and $g$ modes propagating in the radiative core. This coupling occurs when modes of the same degree with similar frequencies interact between cavities, allowing energy to tunnel through the evanescent region that separates them. To model this interaction, the asymptotic framework treats the mixed-mode frequencies as functions of the independent, uncoupled solutions of both the acoustic and gravity cavities. In this section we focus on dipole modes, since these are most important for the analysis.

   For the acoustic cavity, pure $p$ modes such as the radial ($\ell=0$) modes follow an approximately regular frequency pattern \citep[][]{Tassoul1980}. Their frequencies can be described by the so-called universal pattern of solar-like oscillations \citep{Mosser11UP}:
   \begin{equation}\label{eq:UP}
      \nu_{n_{p},\ell} = \left(n_{p} + \frac{\ell}{2} - d_{0\ell} + \epsilon_{p} + \frac{\alpha}{2} (n_{p} - n_{\rm max})^{2}\right)\Delta\nu ,
   \end{equation}
   where $n_{p}$ is the acoustic radial order, $d_{0\ell}$ is the small frequency separation, $\epsilon_{p}$ is a phase offset, and $\alpha$ accounts for the curvature of the $p$-mode pattern. The quantity $n_{\rm max} = \nu_{\rm max}/\Delta\nu$ corresponds to the radial order of the mode closest to $\nu_{\rm max}$. In this work, we assumed that pure dipole $p$ modes ($\nu_{p, 1}$) also follow this universal pattern, with the small separation $d_{01}$ being the offset of the dipole mode frequency from the midpoint between the radial modes.

   \begin{figure*}[t]
      \centering
      \includegraphics[width=0.99\textwidth]{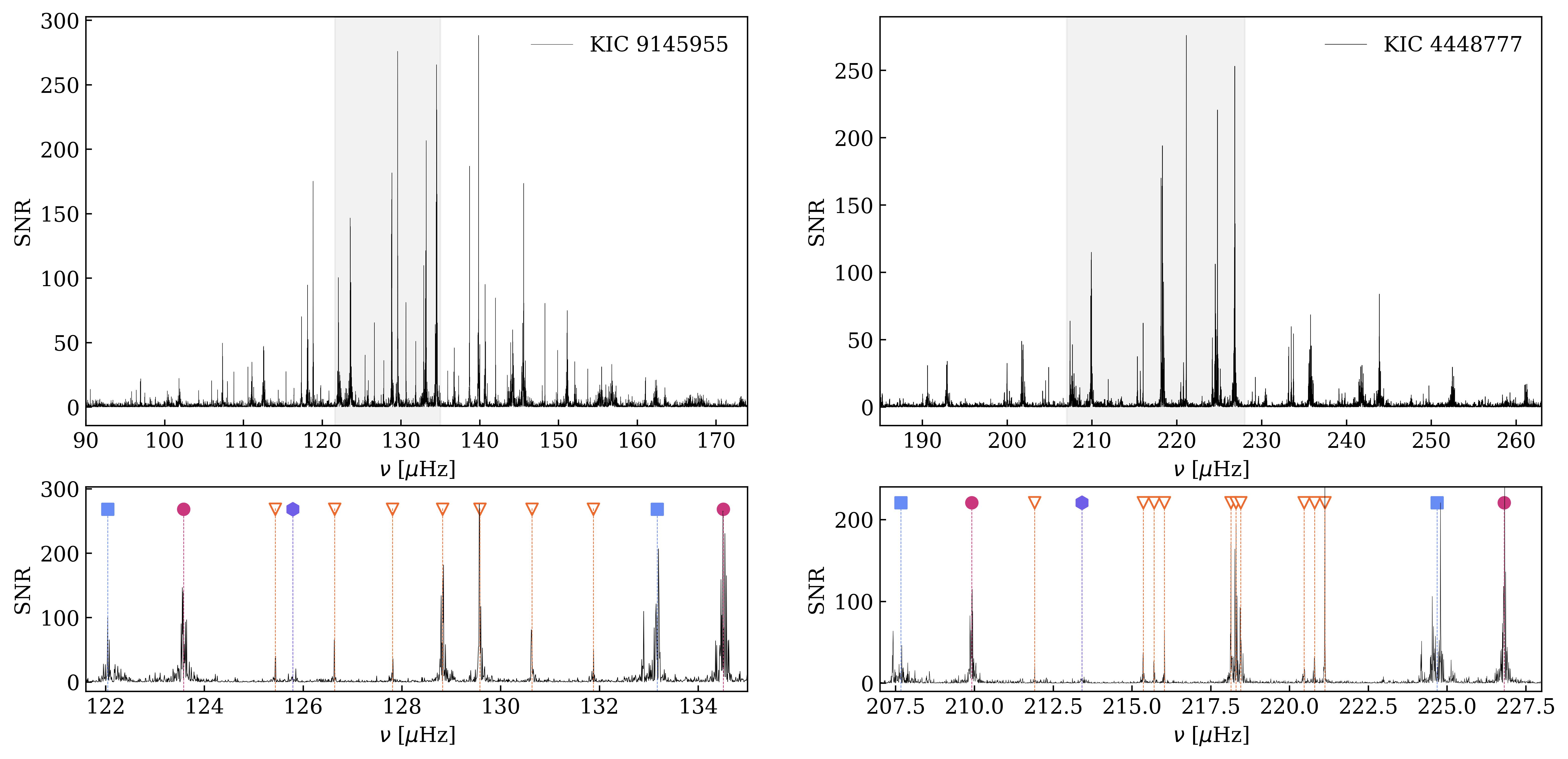} 
      \caption{Power density spectra of the RGB stars KIC~9145955 (left) and KIC~4448777 (right). The lower panels show a zoomed-in view of the central radial orders (grey area in the top panels) for a spectrum with no rotational splittings detected (left), and one where rotational splittings are detected (right). Radial, quadrupole, and octupole modes are indicated as pink circles, blue squares, and purple hexagons. Dipole-mode candidates are depicted with orange triangles.}\label{fig:pds_examplars}
   \end{figure*}

   The frequencies of pure $g$ modes at high radial orders in the gravity cavity satisfy an asymptotic eigenvalue equation \citep[e.g.][]{Aerts10, Lindsay22}:
   \begin{equation}
      \frac{1}{\nu_{g}} = \Delta\Pi_{1}  \left(n_{g} + \frac{1}{2} + \epsilon_{g}\right),
   \end{equation}
   where $\Delta\Pi_{1}$ is the asymptotic period spacing of dipole $g$ modes, $n_{g}$ is the $g$-mode radial order, and $\epsilon_{g}$ is the gravity-phase offset. 
   
   The interaction between these two cavities is commonly described using the implicit asymptotic formalism developed by \citet{Shibahashi79} and \citet{Unno89}, where the frequencies of mixed $\ell=1$ modes are roots of the following equation:
   \begin{equation}\label{eq:asympt_eq}
      \tan \Theta_{p} = q \tan \Theta_{g};
   \end{equation}
   here, $q$ is the coupling factor between the pressure and gravity cavities, and $\Theta_{p}$ and $\Theta_{g}$ are phase terms associated with the pure $p$- and $g$-mode solutions, respectively:
   \begin{align}
      \Theta_{p} &= \pi \frac{\nu - \nu_{n_{p},1}}{\Delta\nu}, \\
      \Theta_{g} &= \pi \frac{1}{\Delta\Pi_{1}} \left(\frac{1}{\nu} - \frac{1}{\nu_{g}}\right).
   \end{align}

   In many red giants, the effects of rotation are detectable through the rotational splittings of these dipole mixed modes \citep[e.g.][]{Beck12,Deheuvels12,DiMauro16}. Stellar rotation lifts the degeneracy in the azimuthal order $m$, producing rotationally split multiplets. Whether these multiplets are resolved depends on the rotation rate relative to the intrinsic mode line widths and frequency resolution. The relative visibility of the azimuthal components is determined by the mode amplitudes and the inclination angle $i$ between the star's rotation axis and the line of sight \citep[e.g.][]{Gizon03}. Hence, dipole mixed modes may appear as singlets $(m=0)$, doublets $(m=\pm1)$, or triplets $(m=0,\pm1)$ in the PDS. Observationally, these morphologies typically correspond to $10^{\circ}<i<20^{\circ}$, $70^{\circ}<i<80^{\circ}$, and $20^{\circ}<i<70^{\circ}$, respectively \citep[e.g.][]{Kamiaka18,Gehan21}. Figure~\ref{fig:pds_examplars} shows the PDS of a star with singlets (KIC~9145955) and a star where the dipole modes are clearly resolved into triplets (KIC~4448777).

   The observed rotational splitting is an average measure of the internal rotation rate, with a mode-dependent sensitivity function acting as a weight \citep[see][]{Ahlborn20}. Consequently, rotational splittings vary with frequency as individual mixed modes probe different regions of the stellar interior. Therefore, they are determined by the rotation rate of the region to which a given mode is most sensitive. \citet{Mosser12b} proposed an empirical prescription to describe this behaviour, where the rotational splitting of dipole mixed modes can be expressed through the implicit relation:
   \begin{align}\label{eq:splittings}
      \nu_{n,1,m} &= \nu_{n,1} + m \, \mathcal{R}_{n_{p}}(\nu_{n,1,m}) \, \delta\nu_{\rm rot}, \\
      \mathcal{R}_{n_{p}}(\nu) &= 1 - \frac{\lambda}{1 + \left( \dfrac{\nu - \nu_{n_{p},1}}{\beta \Delta\nu} \right)^{2}},
   \end{align}
   where $\nu_{n, 1}$ is the frequency of the unperturbed dipole mixed mode, $\delta\nu_{\rm rot}$ is the rotational splitting of the core, and $\mathcal{R}_{n_{p}}$ modulates its amplitude across an acoustic radial order. Because Eq.~(\ref{eq:splittings}) is implicit, the prograde $(m = +1)$ and retrograde $(m = -1)$ components are evaluated at different frequencies along the modulation function $\mathcal{R}$, which introduces an asymmetry within the multiplet. In this work, we adopted this implicit formulation to model the rotational splittings. While this approach accounts for the intrinsic rotational asymmetry, it does not include formalisms for other physical phenomena that can perturb the multiplet symmetry, such as magnetic fields in the stellar core \citep[e.g.][]{Fuller15,LiGang23,Mathis23,Hatt24}, near-degeneracies \citep[e.g.][]{Deheuvels17,Ahlborn25}, or second-order rotational effects \citep[e.g.][]{Dziembowski92,Ouazzani13}. Therefore, any asymmetries detected by our algorithm cannot be straightforwardly attributed to a single physical origin without more detailed investigation.

   Together, Eqs.~(\ref{eq:UP}) to (\ref{eq:splittings}) represent the set of equations we used to compute dipole mixed-mode frequencies in our pipeline. From here on, we refer to this set as the ``mixed-mode equation''. We used $\lambda=0.5$ and $\beta=0.08$ following \citet{Mosser12b}, who found these values to remain consistent across different stars and different evolutionary states. Therefore, our pipeline treated $\nu_{\rm max}$, $\Delta\nu$, $\epsilon_{p}$, $\alpha$, $d_{01}$, $\Delta\Pi_{1}$, $q$, $\epsilon_{g}$, and $\delta\nu_{\rm rot}$ as free parameters.

\section{Frequency-template overlap score}\label{sec:score_description}

   We compared observed and theoretical dipole mixed-mode frequencies by representing them as continuous templates rather than discrete lists. Specifically, we modelled each frequency set as a spike train convolved with unit-height Gaussian kernels of width $\sigma$.
   We then quantified the overlap between observed and theoretical templates using a metric defined by the inner product of the spike trains. By construction, this approach enables a global comparison of frequency sets and does not require one-to-one matching during the optimisation process. Analogous template-based similarity measures have been successfully applied in other fields dealing with sparse point processes\footnote{A sparse point process generates stochastic, discrete events scattered within a bounded spatial, temporal, or frequency domain \citep[e.g.][]{Kass14}.} such as neuroscience \citep[e.g.][]{VictorPurpura96,VictorPurpura97,vanRossum01,Schreiber03}.

   \subsection{Formulation of the overlap score}\label{subsec:score}
      Let $\mathcal{O}$ be a set of $N$ observed frequencies and $\mathcal{P}$ a set of $M$ theoretical frequencies predicted by the mixed-mode equation,
      \begin{align}
      \mathcal{O} = \{\nu_{\mathfrak{o},1}, \nu_{\mathfrak{o},2}, \ldots, \nu_{\mathfrak{o},N}\}, \quad
      \mathcal{P} = \{\nu_{\mathfrak{p},1}(\theta), \nu_{\mathfrak{p},2}(\theta), \ldots, \nu_{\mathfrak{p},M}(\theta)\},
      \end{align}
      where $\theta$ denotes the set of parameters of the mixed-mode equation.
      We represented each frequency set as a spike train,
      \begin{align}
      f_\mathfrak{o}(\nu) = \sum_{i=1}^{N} \delta(\nu - \nu_{\mathfrak{o},i}), \quad
      f_\mathfrak{p}(\nu) = \sum_{j=1}^{M} \delta(\nu - \nu_{\mathfrak{p},j}(\theta)),
      \end{align}
      where $\delta$ is the Dirac delta distribution.
      To construct the frequency templates, we convolved each spike train with a zero-centred, unit-height Gaussian kernel,
      \begin{equation}
      \mathcal{G}(\sigma,\nu) = \exp\!\left[-\frac{\nu^2}{2\sigma^2}\right].
      \end{equation}
      This convolution yields the continuous observed and theoretical frequency templates
      \begin{align}
      \mathcal{T}_\mathfrak{o}(\nu,\sigma_{\mathfrak{o}}) &= \sum_{i=1}^{N} \exp\!\left[-\frac{(\nu-\nu_{\mathfrak{o},i})^2}{2\sigma_{\mathfrak{o}}^2}\right], \\
      \mathcal{T}_\mathfrak{p}(\nu,\sigma_{\mathfrak{p}},\theta) &= \sum_{j=1}^{M} \exp\!\left[-\frac{(\nu-\nu_{\mathfrak{p},j}(\theta))^2}{2\sigma_{\mathfrak{p}}^2}\right].
      \end{align}

      Since $\mathcal{T}_\mathfrak{o}$ and $\mathcal{T}_\mathfrak{p}$ are square-integrable functions belonging to the Hilbert space $L^2(\mathbb{R})$, their inner product is well defined. We therefore defined the ``overlap score'' (or simply ``score'') between the observed and theoretical templates as
      \begin{equation}
         \mathcal{S}_{\mathfrak{o},\mathfrak{p}}(\sigma,\theta) = \langle \mathcal{T}_\mathfrak{o}, \mathcal{T}_\mathfrak{p} \rangle = \int_{-\infty}^{\infty} \mathcal{T}_\mathfrak{o}(\nu,\sigma_{\mathfrak{o}})\, \mathcal{T}_\mathfrak{p}(\nu,\sigma_{\mathfrak{p}},\theta)\, \mathrm{d}\nu.
      \end{equation}
      
      By assuming the same Gaussian width for all observed and predicted modes ($\sigma=\sigma_{\mathfrak{o}}=\sigma_{\mathfrak{p}}$), this integral yields the analytical expression
      \begin{equation}\label{eq:Sx}
         \mathcal{S}_{\mathfrak{o},\mathfrak{p}}(\sigma,\theta) = \sqrt{\pi}\,\sigma \sum_{i=1}^{N} \sum_{j=1}^{M} \exp\!\left[-\frac{(\nu_{\mathfrak{o},i}-\nu_{\mathfrak{p},j}(\theta))^2}{4\sigma^2}\right],
      \end{equation}
      where the factor $\sqrt{\pi}\sigma$ sets the overall scale of the score. 
     
      As defined, $\mathcal{S}_{\mathfrak{o},\mathfrak{p}}(\sigma, \theta)$ is highly sensitive to the number of modes in the observed and theoretical templates. This dependence on the number of modes makes comparing different frequency sets challenging. Furthermore, because unmatched theoretical modes contribute negligibly to the score rather than being penalised, $\mathcal{S}_{\mathfrak{o},\mathfrak{p}}$ alone cannot effectively discriminate against frequency sets with an excessive number of theoretical modes ($M \gg N$). To obtain a scale-invariant measure that accounts for the relative number of modes in the templates, we normalised the score by the geometric mean of the self-overlaps:
      \begin{equation}
         \tilde{\mathcal{S}}(\sigma, \theta) = \frac{\mathcal{S}_{\mathfrak{o},\mathfrak{p}}(\sigma, \theta)}{\sqrt{\mathcal{S}_{\mathfrak{o},\mathfrak{o}}(\sigma)\,\mathcal{S}_{\mathfrak{p},\mathfrak{p}}(\sigma, \theta)}},
      \end{equation}
      where the self-overlaps $\mathcal{S}_{\mathfrak{o},\mathfrak{o}}$ and $\mathcal{S}_{\mathfrak{p},\mathfrak{p}}$ are given by
      \begin{align}\label{eq:Spp_Soo}
         \mathcal{S}_{\mathfrak{o},\mathfrak{o}} &= \int \mathcal{T}_{\mathfrak{o}}^{2}(\nu)\,\mathrm{d}\nu = \sqrt{\pi}\sigma \sum_{i=1}^{N}\sum_{i'=1}^{N}
         \exp\!\left[-\dfrac{(\nu_{\mathfrak{o}, i}-\nu_{\mathfrak{o}, i'})^{2}}{4\sigma^{2}} \right], \\
         \mathcal{S}_{\mathfrak{p},\mathfrak{p}} &= \int \mathcal{T}_{\mathfrak{p}}^{2}(\nu,\theta)\,\mathrm{d}\nu = \sqrt{\pi}\sigma \sum_{j=1}^{M}\sum_{j'=1}^{M}
         \exp\!\left[-\dfrac{(\nu_{\mathfrak{p}, j}-\nu_{\mathfrak{p}, j'})^{2}}{4\sigma^{2}} \right].
      \end{align}

      This normalisation corresponds to the cosine of the angle between the observed and theoretical templates in the $L^2(\mathbb{R})$ Hilbert space, and it is commonly referred to as ``cosine similarity'' \citep[e.g.][]{Manning08}.
      By construction, $\tilde{\mathcal{S}}$ is symmetric, invariant under uniform rescaling of the templates, and bounded between 0 and 1. A perfect match yields $\tilde{\mathcal{S}}=1$, while a total lack of overlap results in $\tilde{\mathcal{S}}=0$. Importantly, normalising by the self-overlaps ensures that an excessive number of modes in either the observed or the theoretical templates, relative to the other, results in a lower overlap score.
      
   \subsection{Gaussian kernel width bounds}\label{subsec:sigma}

      \begin{figure}[t]
         \centering
         \includegraphics[width=0.49\textwidth]{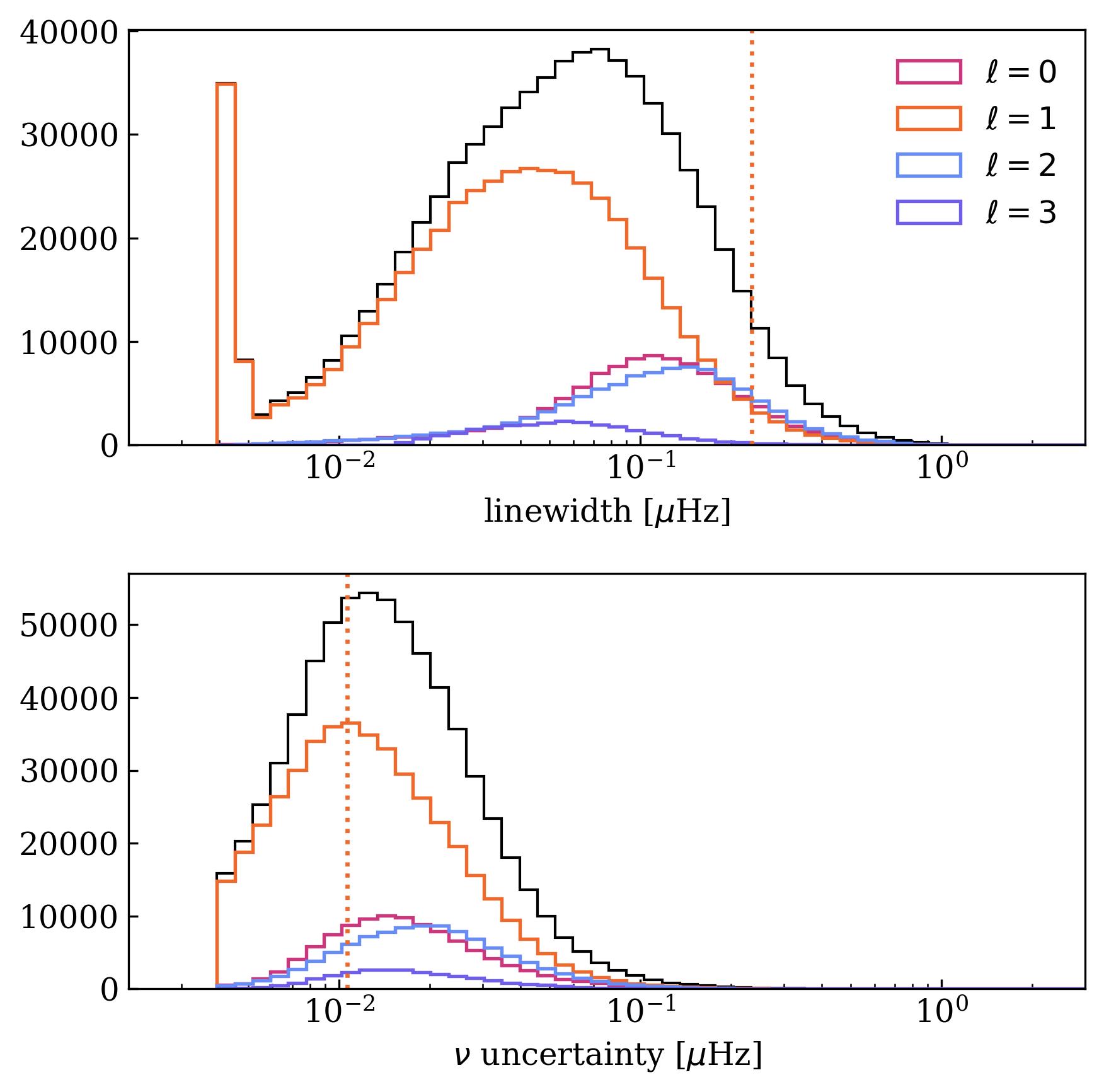} 
         \caption{Top: Distribution of mode line widths for different spherical degrees. Bottom: Distribution of mode frequency uncertainties for different spherical degrees. The vertical dashed lines show the 95th percentile of the line-width distribution and the 50th percentile of the frequency uncertainty distribution, respectively. The black line shows the cumulative distribution for all $\ell$.}\label{fig:sigma}
      \end{figure}

      \begin{figure*}[t]
         \centering
         \includegraphics[width=0.94\linewidth]{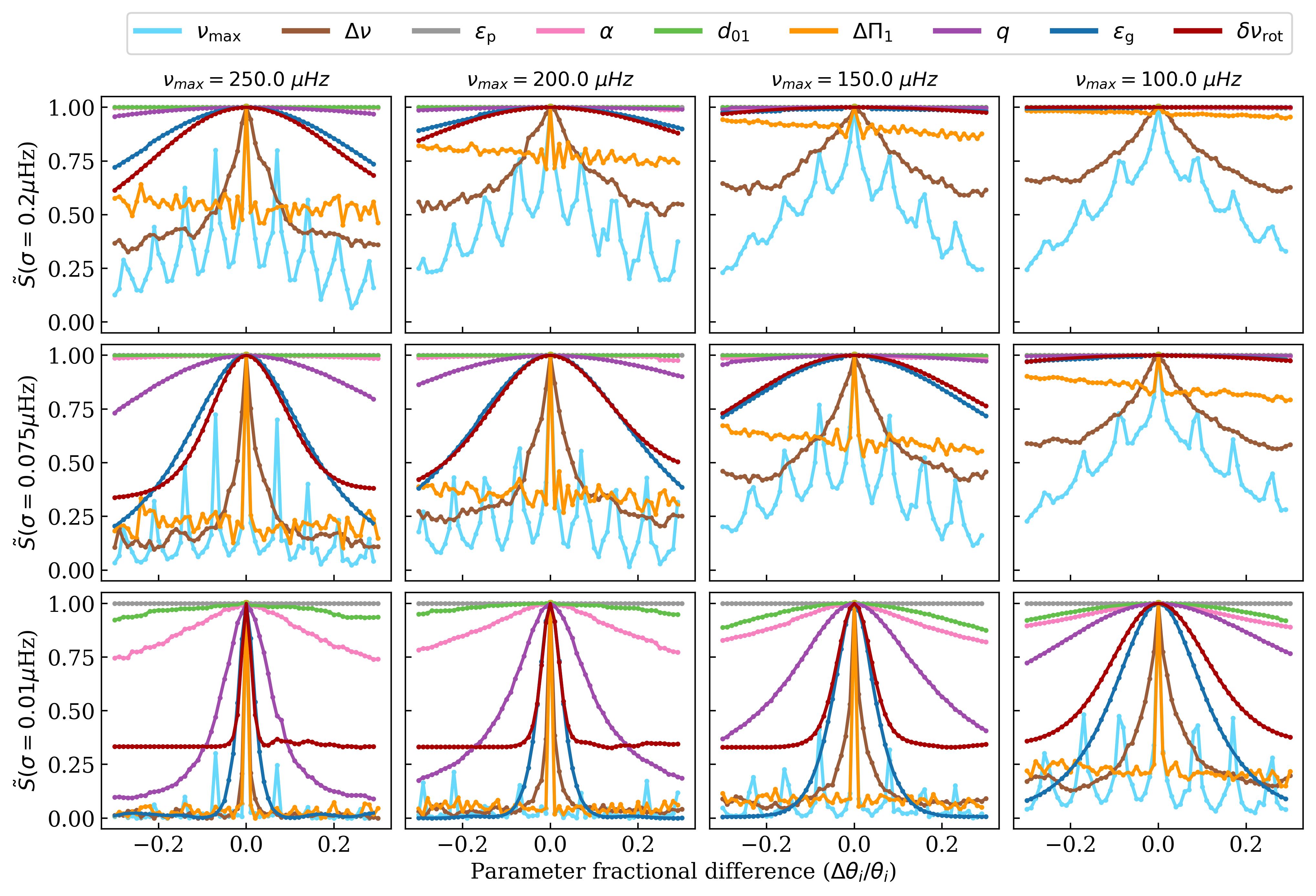}
         \caption{Response profiles of the normalised score $\tilde{\mathcal{S}}(\sigma, \theta)$ to fractional perturbations of the parameters of the mixed-mode equation. Each column represents a synthetic RGB star at a different $\nu_{\rm max}$ ($100, 150, 200,$ and $250$~$\mu$Hz). Rows show the results for three Gaussian kernel widths: $\sigma=0.2, 0.075,$ and $0.01$~$\mu$Hz. Each colour corresponds to a different parameter, with the central peak ($\Delta\theta_{i}/\theta_{i}=0$) indicating the unperturbed value.}\label{fig:mbw_score}
      \end{figure*}

      Given that the score strongly depends on $\sigma$, its bounds must be chosen so that they are wide enough to account for the physical resolution of the peaks, yet narrow enough to prevent adjacent modes from blending. Since this optimal trade-off depends on the length of the observations (which determines the frequency resolution), the global bounds must be calibrated to the specific dataset. In this work, we established empirical bounds for the $\sim4$-year \emph{Kepler} data using a sample of approximately 15,600 red giants. These were analysed with TACO (Tools for the Automated Characterisation of Oscillations; Hekker et al., in prep.)\footnote{We refer the reader to the works of \citet{Themessl18a}, \citet{Choi25}, and \citet{FER25} for a detailed description of TACO.}, and are reported in CAPASS (Catalogue of Asteroseismic Parameters for Analysis of Stellar Structure; Espinoza-Rojas et al., in prep.). Figure~\ref{fig:sigma} illustrates the observed distributions of mode line widths and frequency uncertainties. 
      
      We set the global upper bound to $\sigma\sim0.2\,\mu$Hz, which corresponds to the 95th percentile of the line-width distribution for dipole mode candidates (i.e. modes not identified by TACO as radial, quadrupole, or octupole). For the lower bound, we adopted the 50th percentile of the frequency uncertainty distribution of dipole modes, approximately $0.01\,\mu$Hz. 
      
      These values define the plausible range of $\sigma$ for the nominal \emph{Kepler} mission. We emphasise that when applying this method to data from other missions, users must define new bounds for $\sigma$ to account for different frequency resolutions. However, this recalibration can be done by following the same logic presented here.

   \subsection{Response profiles of the overlap score}\label{subsec:score_sensitivity}
      
      To assess how the normalised score $\tilde{\mathcal{S}}$ responds to variations in the parameters of the mixed-mode equation, we evaluated the score response profile for each parameter $\theta_{i}$. We characterised the sensitivity of the metric as the gradient of this profile, i.e. $\partial \tilde{\mathcal{S}}(\sigma,\theta)/\partial \theta_i$. A narrow, prominent peak indicates that $\tilde{\mathcal{S}}$ is highly sensitive to that parameter, as the overlap is maximised only within a tight range around the reference value. This analysis quantifies how strongly the dipole mixed-mode pattern depends on each parameter and how this dependence varies with the choice of kernel width $\sigma$. 

      We generated synthetic dipole frequencies of RGB stars at $\nu_{\rm max}=100$, 150, 200, and $250\,\mu$Hz with triplets using the mixed-mode equation, with parameters derived from the scaling relations in Appendix~\ref{appendix:scaling_relations}. Throughout this work, we refer to these theoretical frequency sets as ``synthetic stars''. We then applied a relative perturbation of up to $\pm30\%$ to each parameter in steps of $1\%$ while holding the others fixed. For each perturbation, we recomputed the normalised score between the reference and perturbed templates.   
      
      Figure~\ref{fig:mbw_score} shows the resulting response profiles of the normalised score for $\sigma=0.2, 0.075,$ and $0.01\,\mu$Hz. We find that the score is most sensitive to the period spacing ($\Delta\Pi_{1}$), as perturbations of $\sim1\%$ produce a sharp decrease in $\tilde{\mathcal{S}}$. This is expected, as the period spacing determines the relative alignment of the entire frequency pattern.
      In contrast, the normalised score exhibits negligible sensitivity to $\epsilon_{p}$. Sensitivity to parameters such as $\epsilon_{g}$, the curvature $\alpha$, and the coupling factor $q$ only becomes significant when $\sigma$ approaches 0.01~$\mu$Hz. The multiple local peaks in $\nu_{\rm max}$ correspond to variations that are integer multiples of $\Delta\nu$, producing the observed comb-like pattern, which becomes less prominent for narrower $\sigma$. Additionally, the sensitivity of $\tilde{\mathcal{S}}$ to $\delta\nu_{\rm rot}$ is similar to its sensitivity to $\epsilon_{g}$. However, for $\delta\nu_{\rm rot}$, $\tilde{\mathcal{S}}$ plateaus at around 0.33 when deviations in $\delta\nu_{\rm rot}$ become too large, as only the $m=0$ mode remains aligned between the templates.

      The behaviour of the parameters in Fig.~\ref{fig:mbw_score} illustrates an important trade-off. Smaller values of $\sigma$ lead to narrower frequency templates and enhance the ability to resolve subtle shifts; however, this comes at the cost of a smaller frequency tolerance around each mode, restricting the allowed margin of error between the theoretical and observed frequencies. A very small $\sigma$ reduces the volume of the parameter space where the overlap is significant, potentially creating a score surface (i.e. the map of scores across all possible parameter combinations) dominated by narrow peaks, which complicates locating the best-fit parameters. Conversely, larger $\sigma$ values capture global features of the mixed-mode pattern even if individual frequencies are not perfectly aligned. Hence, the score incorporates an explicit frequency tolerance by using Gaussian kernels of finite width. Finally, Fig.~\ref{fig:mbw_score} shows that the sensitivity at a fixed $\sigma$ decreases as stars ascend the RGB, and therefore, $\sigma$ must be defined on a star-by-star basis.

   \subsection{Multi-bandwidth formulation}\label{subsec:Smbw}
      The trade-off between kernel width and parameter sensitivity discussed in Sect.~\ref{subsec:score_sensitivity} suggests that a single value of $\sigma$ is typically insufficient to characterise the full mixed-mode pattern. To address this, we combined the advantages of broad and narrow kernels into a single multi-bandwidth score,
       \begin{equation}
         \tilde{\mathcal{S}}_{\rm mbw}(\hat{\sigma},\theta) = \frac{1}{K}\sum_{k=1}^{K}  \tilde{\mathcal{S}}(\sigma_{k}, \theta),
      \end{equation}
      where $\hat{\sigma} = \{\sigma_{k}\}$ is a set of Gaussian widths defined for an individual star. By combining these scales, the resulting score surface is sensitive to all parameters of the mixed-mode equation. Moreover, it adapts the metric to the intrinsic variation in dipole-mode line widths while avoiding the mode blending that would occur if only a single broad kernel were used.
     
\section{Optimisation and mode identification}\label{sec:optimisation_modeID}
   To find the parameters of the mixed-mode equation that provide the highest overlap, we maximised $\tilde{\mathcal{S}}_{\rm mbw}$ by performing a global optimisation. Because some parameters in the mixed-mode equation are degenerate, the score surface can be highly non-convex with numerous local maxima. This makes traditional gradient-based solvers and standard sampling techniques (such as Markov chain Monte Carlo approaches) prone to premature convergence. We therefore employed a genetic algorithm (GA), a population-based method capable of exploring highly degenerate parameter spaces to find the global maximum. This method simultaneously tests a diverse set of possible solutions and introduces random variations. Hence, a GA can broadly search the parameter space without becoming trapped in these local maxima. We used the \texttt{pyGAD} implementation \citep{pygad}. Details regarding the algorithm configuration and hyperparameters are provided in Appendix~\ref{appendix:GA}.

   To compute the full set of theoretical dipole mode frequencies per star, we employed a two-step numerical root-finding procedure. First, we computed the unperturbed dipole mixed modes (i.e. $m=0$) on a radial-order basis, finding the roots of Eq.~(\ref{eq:asympt_eq}) with a fixed $\theta$ across all orders. To do so, we evaluated the equation on a frequency grid and identified the roots as points where the function changes sign. By restricting the search to the expected frequency range of dipole modes for a given order, we limited the required frequency grid size and significantly reduced the computational overhead. 
   
   Second, we computed the rotationally split modes (i.e. $m=\pm1$). Since the frequency shift depends on the frequency of the mode itself (Eq.~\ref{eq:splittings}), the relation is implicit. We therefore solved it using the Newton--Raphson method. We calculated the exact analytical derivative of the rotation profile and initialised the solver with an estimate of the rotational shift evaluated at the unperturbed frequency. We then varied the frequency until the difference between iterations dropped below a tolerance level of $10^{-6}~\mu$Hz.

   To account for the effects of stellar inclination on the morphology of rotationally split multiplets, we performed the optimisation for three different multiplet types (i.e. singlets, doublets, and triplets) in parallel. During each optimisation run, the multiplet type remained fixed while the parameters of the mixed-mode equation were allowed to vary. We first identified the best-fitting set of parameters by selecting the configuration with the highest final score. As a final refinement, we took the best-fit parameters and re-evaluated the score across all three multiplet types while keeping the parameters fixed. This two-step approach ensures that the selected multiplet type is the one that truly maximises the overlap, allowing a proper identification in cases where distinguishing between doublets and triplets is challenging, for example when the visibility of the $m=0$ component is low and therefore not always detectable in all the observed dipole modes.

   Once the optimal parameters were determined, we performed the automated identification of dipole mixed modes. We first computed the theoretical frequencies from the best-fit parameters. We then found the optimal assignment between the observed frequencies and the theoretical template using the Hungarian method \citep{Kuhn55}, as implemented in the \texttt{scipy}\footnote{\url{https://docs.scipy.org/doc/scipy/index.html}} module. To ensure physical consistency, we introduced a distance threshold in the cost matrix so that any potential assignments where the frequency difference exceeds $3\sigma_{\rm max}$ ($\sigma_{\rm max}$ being the largest Gaussian width in the multi-bandwidth set) were discarded. This prevents the algorithm from forcing a match between modes that are outside the frequency tolerance.

   \subsection{Computational performance}
      \begin{figure}
         \centering
         \includegraphics[width=0.45\textwidth]{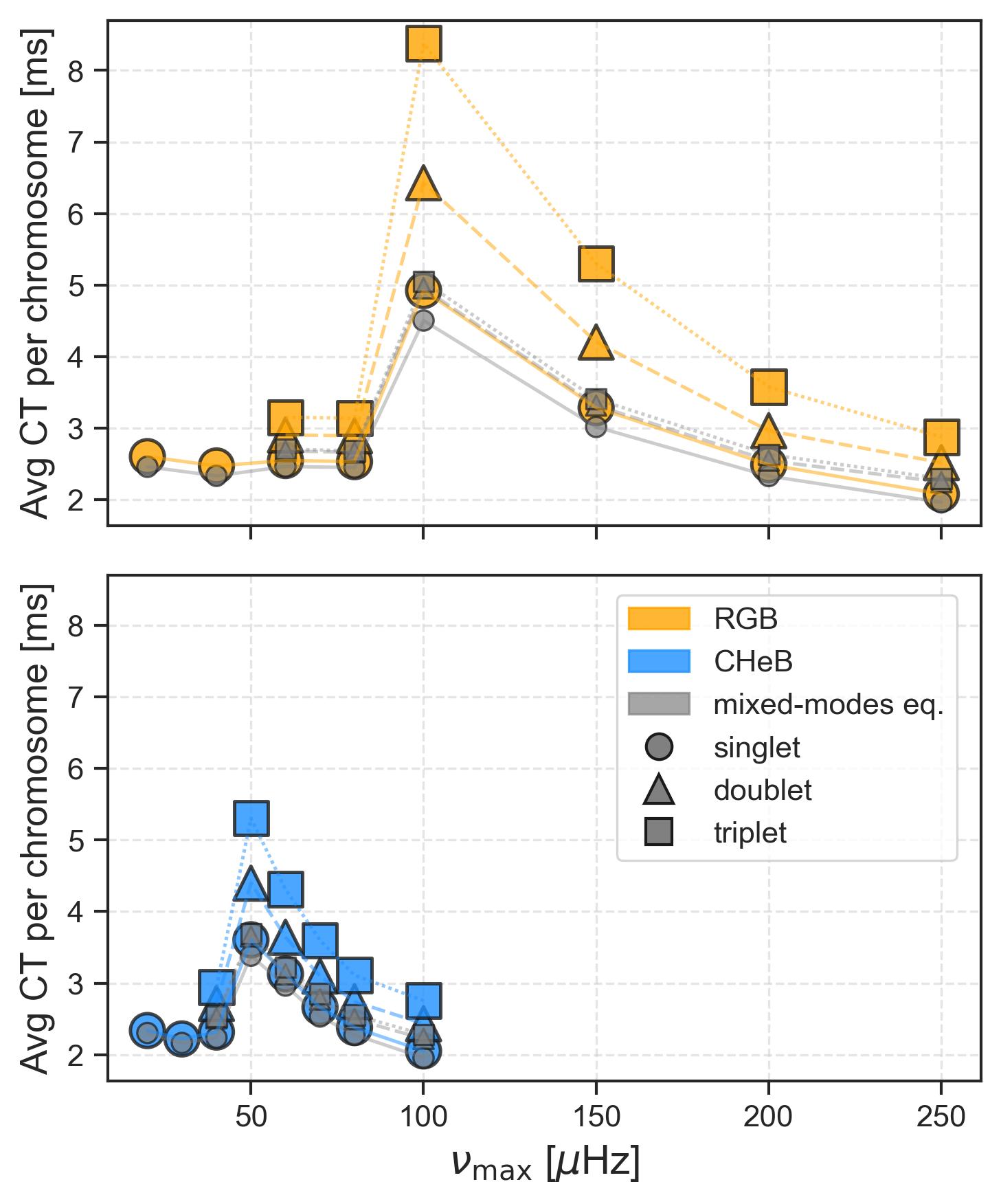}
         \caption{Average computing time (CT) per chromosome as a function of $\nu_{\rm max}$. RGB and CHeB stars are shown in orange and blue, respectively. Synthetic stars with singlets, doublets, and triplets are indicated with circles, triangles, and squares, respectively. The CT for solving the mixed-mode equation is shown in grey.}\label{fig:CT}
      \end{figure}
      Our pipeline repeatedly evaluates the mixed-mode equation, performing numerical root-finding across many chromosomes and generations in the GA (see Appendix~\ref{appendix:GA}). To avoid a computational bottleneck during the global optimisation, we restricted the root-finding grid search to specific frequency intervals for each radial order. Furthermore, we accelerated the calculations using the {\tt Numba} just-in-time compilation package\footnote{\url{http://numba.pydata.org}}.
      
      We assessed the algorithm's overall performance on an Apple M1 Pro chip using synthetic stars. As illustrated in Fig.~\ref{fig:CT}, the average CT per chromosome depends primarily on the evolutionary state, $\nu_{\rm max}$, and the multiplet type. The CT increases as stars evolve towards lower $\nu_{\rm max}$ because of the higher density of dipole mixed modes. The sharp decrease in CT below $100~\mu$Hz for RGB stars and $50~\mu$Hz for CHeB stars is a natural consequence of how the synthetic stars were generated, as we trimmed dipole modes to mimic observed dipole-mode densities (see Sect.~\ref{subsec:grid}). As expected, models with triplets take longer to compute than those with doublets, and the same holds for doublets compared with singlets. Overall, optimising CHeB synthetic stars is typically faster than optimising RGB stars. The same performance is observed when tested on a multicore setup (e.g. using six CPU cores on an Intel Xeon Gold 6230 cluster).

      The total runtime per star scales directly with the user-defined population size (i.e. the number of chromosomes per generation) and the convergence criteria (see Appendix~\ref{appendix:GA} for definitions). Under the configuration used in this work, the GA optimisation typically converges within 30 minutes to a few hours per star. Hence, the numerical treatment of the implicit equations is computationally efficient enough for large-scale automated analyses.
          
\section{Validation on synthetic stars}\label{sec:validation}
   We validated the proposed method using a grid of theoretical dipole mode frequency sets. This grid allowed us to verify whether the GA can accurately recover the parameters of the mixed-mode equation when the true values are known a priori. The synthetic stars in the grid were constructed using the same methodology described in Sect.~\ref{subsec:score_sensitivity}, with dipole-mode frequencies spanning seven acoustic orders\footnote{We typically observe around seven acoustic orders in \emph{Kepler} data.}. To ensure the method is robust across different evolutionary states, our grid includes synthetic RGB stars with $\nu_{\rm max}$ between $20$ and $250$\,$\mu$Hz, and CHeB stars between $20$ and $100$\,$\mu$Hz. To simulate observational noise, we then perturbed the theoretical frequencies by adding Gaussian noise with a standard deviation of $0.01$\,$\mu$Hz, which is approximately the typical frequency resolution of the \emph{Kepler} data. We defined $\sigma$ per synthetic star (see Table~\ref{table:sigma_list}), with upper and lower bounds set by the global limits found in Sect.~\ref{subsec:sigma} ($\sim0.01$ to $0.2~\mu$Hz). To mimic observations, we empirically scaled the subset of $\sigma$ values for each synthetic star as a function of its $\nu_{\rm max}$ and evolutionary state.
   
   \subsection{Construction of the synthetic grid}\label{subsec:grid}
\begin{figure}[t]
         \centering
         \includegraphics[width=0.48\textwidth]{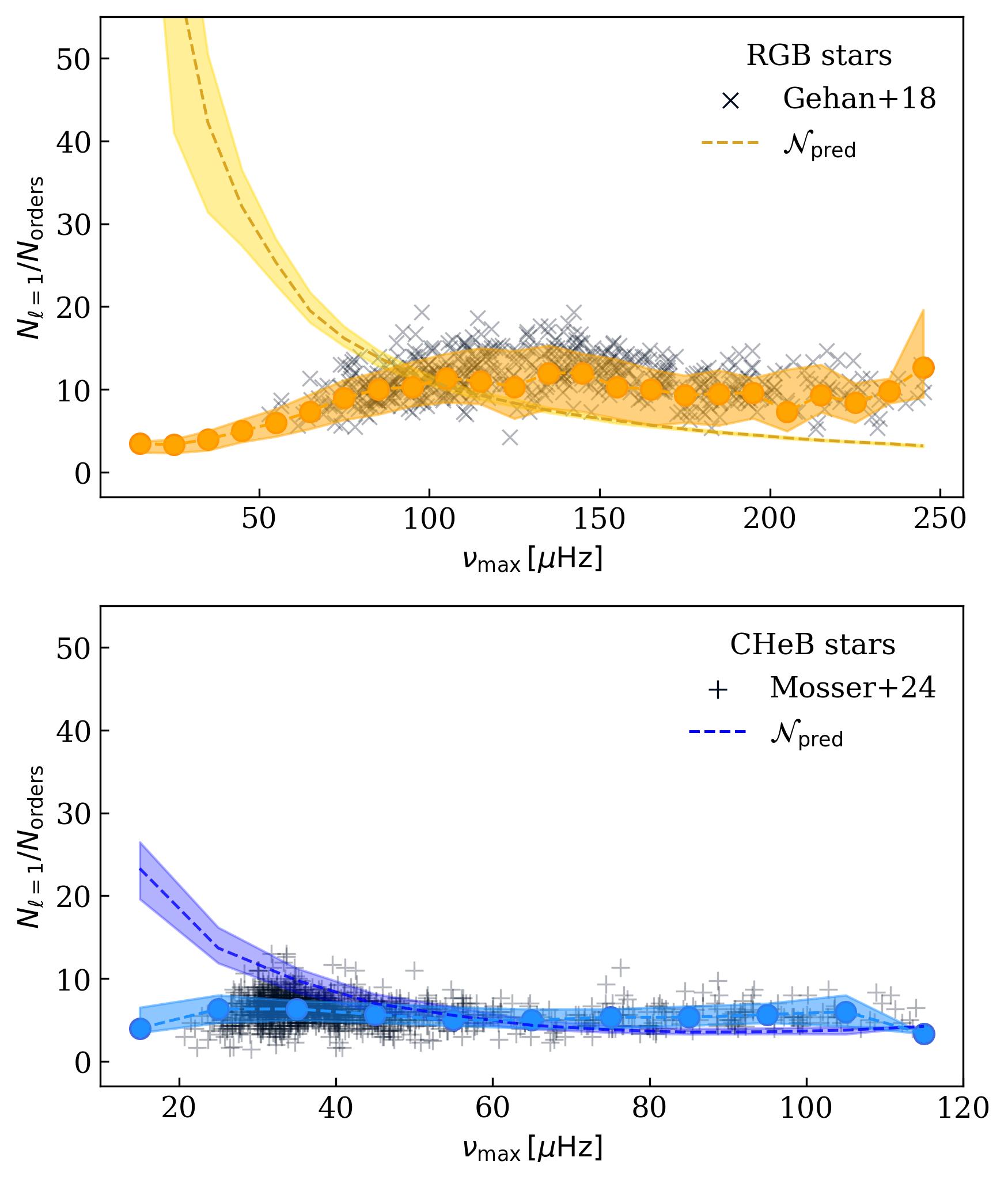}
         \caption{Comparison of the observed and expected dipole mixed-mode density $(\mathcal{N})$ as a function of $\nu_{\rm max}$. Top: RGB stars. Bottom: CHeB stars. Orange and blue dots represent the median observed mode density for RGB and CHeB stars, respectively. The dashed lines indicate the median theoretical density. Shaded regions around the lines and dots correspond to the 16th and 84th percentiles of the respective distributions within each frequency bin. Crosses and plus symbols are stars with rotation rates from \citet{Gehan18} and \citet{Mosser24}, respectively.}\label{fig:l1_density}
      \end{figure}

      To ensure our synthetic grid is as realistic as possible, it must reflect the number of mixed modes that are typically detectable in observations at different evolutionary states. Accordingly, we defined the observed mode density around $\nu_{\rm max}$ as the ratio between the number of dipole-mode candidates and the number of acoustic radial orders within a frequency range of $\nu_{\rm max} \pm 2\Delta\nu$. We compared this empirical estimate with the theoretical density of dipole mixed modes within a radial order, which we assumed to be approximated as $\mathcal{N} = \Delta\nu^{2} / (\Delta\Pi_{1} \nu_{\rm max}^{2})$ \citep[e.g.][]{LiGang24}. The median theoretical density shown in Fig.~\ref{fig:l1_density} represents the median of the individual $\mathcal{N}$ values, which were calculated for each star using its observed $\nu_{\rm max}$, $\Delta\nu$, and $\Delta\Pi_{1}$ reported in the CAPASS catalogue. 

      \begin{figure*}[!t]
         \centering
         \includegraphics[width=0.98\textwidth]{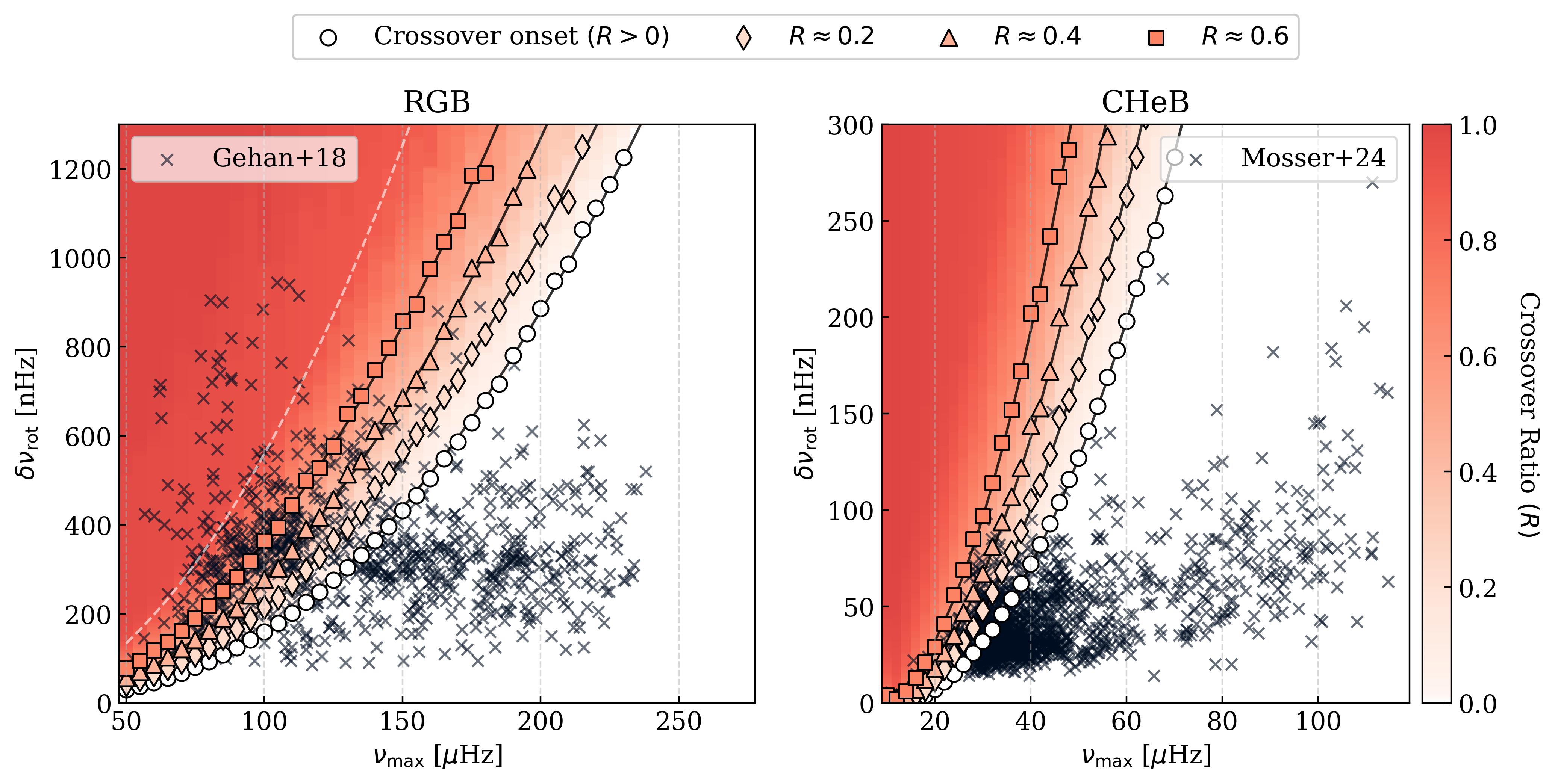}
         \caption{Crossover ratio, $R$, as a function of $\nu_{\rm max}$ and $\delta\nu_{\rm rot}$. The background is colour-coded by the crossover ratio. The markers indicate specific grid points for the crossover onset (circles), $R=0.2$ (diamonds), $0.4$ (triangles), and $0.6$ (squares). The dashed white line indicates $R=0.9$ for RGB stars. Black lines show the second-order polynomial fit for each of these crossover levels. Crosses represent rotational splittings reported in the literature: \citet{Gehan18} for RGB stars (left panel) and \citet{Mosser24} for CHeB stars (right panel).}  \label{fig:rot_regimes}
      \end{figure*}

      As illustrated in Fig.~\ref{fig:l1_density}, the observed dipole-mode density typically exceeds theoretical values at higher frequencies, specifically for RGB stars with $\nu_{\rm max}>100$~$\mu$Hz and CHeB stars with $\nu_{\rm max}>50$~$\mu$Hz. This over-density is explained by the presence of dipole-mode multiplets, as stars with rotational splittings from \citet{Gehan18} (RGB) and \citet{Mosser24} (CHeB) consistently fall above the theoretical threshold. In contrast, observed densities at lower frequencies are significantly lower than theoretical predictions. For RGB stars with $\nu_{\rm max} < 100$~$\mu$Hz, the coupling between $p$ and $g$ modes is relatively weak, meaning that only the most $p$-dominated mixed modes have sufficient amplitudes to be detected above the noise. Conversely, in the regime with $\nu_{\rm max}< 40$~$\mu$Hz for CHeB stars, the strong coupling \citep[$q \approx 0.3$; e.g.][]{Vrard16,Mosser24}, in combination with the intrinsically large $g$-mode inertias, results in mixed modes with small amplitudes due to the dilution of the energy of $p$ modes. This interaction produces mixed-mode amplitudes falling below the detectability threshold.

      To reproduce this behaviour of RGB and CHeB stars at lower frequencies, we adjusted our synthetic stars to match the observed dipole-mode density. Specifically, for synthetic RGB stars with $\nu_{\rm max}<100~\mu$Hz and CHeB stars with $\nu_{\rm max}<50$~$\mu$Hz, we retained only the four most $p$-dominated $\ell=1$ modes per acoustic order. By selecting the modes themselves rather than individual frequencies, we ensured that the entire rotational multiplet (whether a singlet, doublet, or triplet) was preserved for each mode.

      Additionally, to test the method's performance at different core rotation rates, we included rotational splittings for all CHeB synthetic stars and for RGB stars with $\nu_{\rm max} \geq 50~\mu$Hz. Below this threshold, the extremely weak coupling traps $g$-dominated modes in the core, making the detection of rotational splittings unlikely in \emph{Kepler} data \citep[e.g.][]{Mosser18}. We set the rotational splitting of each synthetic star based on a proxy for the complexity of the frequency pattern due to rotation. This complexity is defined by the number of peak crossovers. A crossover occurs when the prograde $(m=+1)$ component of a mode $i$ shifts past the retrograde $(m=-1)$ component of the adjacent mode $i+1$ (i.e. $\nu_{i, m=+1}>\nu_{i+1, m=-1}$). We defined the ``crossover ratio'', $R$, as the total number of crossovers divided by the total number of adjacent mixed modes. A value of $R=0$ indicates completely separated multiplets, whereas higher values correspond to increasing pattern complexity. 

      Because the mixed-mode density changes strongly with evolution, a constant rotational splitting produces different degrees of mode crossing at different $\nu_{\rm max}$. To define consistent crossover levels, we computed triplet frequencies on a grid in $\nu_{\rm max}$ and $\delta\nu_{\rm rot}$ (up to $1300~$nHz for RGB and $500~$nHz for CHeB stars), and evaluated $R$ at each grid point (see Fig.~\ref{fig:rot_regimes}). We then identified the splitting values that produce four specific regimes: a baseline case, $R\approx 0.0$, corresponding to the maximum splitting with no crossovers, and $R\approx 0.2, 0.4$, and $0.6$, where 20\%, 40\%, and 60\% of adjacent multiplets intersect, respectively. Finally, we fitted these $\delta\nu_{\rm rot}$ values as a function of $\nu_{\rm max}$ using a second-order polynomial. The polynomial coefficients for each $R$ are provided in Table~\ref{tab:rotation_coeffs}. We then used these fits to assign $\delta\nu_{\rm rot}$ for the synthetic grid. 

      Interestingly, there is a subset of RGB stars from \citet{Gehan18} in Fig.~\ref{fig:rot_regimes} with crossover ratios larger than $\sim0.9$. After visual inspection of these targets, we find that their PDS exhibit a low number of observed modes, which complicates the unambiguous measurement of such large splittings. Furthermore, in stars where splittings are visible, they can often be explained by slower core rotation in combination with rotational asymmetries. However, we did not perform any detailed analysis of these stars, as it is outside the scope of this work. Therefore, to ensure a conservative range for $\delta\nu_{\rm rot}$, we adopted $R\approx0.6$ as a representative upper limit for the typical red-giant rotational splittings.
      
      In total, our grid comprises 128 unique synthetic stars featuring different multiplet types (i.e. singlets, doublets, and triplets) covering the four crossover levels ($R\approx0.0, 0.2, 0.4,$ and $0.6$).

      \begin{figure*}[ht]
         \centering
         \includegraphics[width=0.98\textwidth]{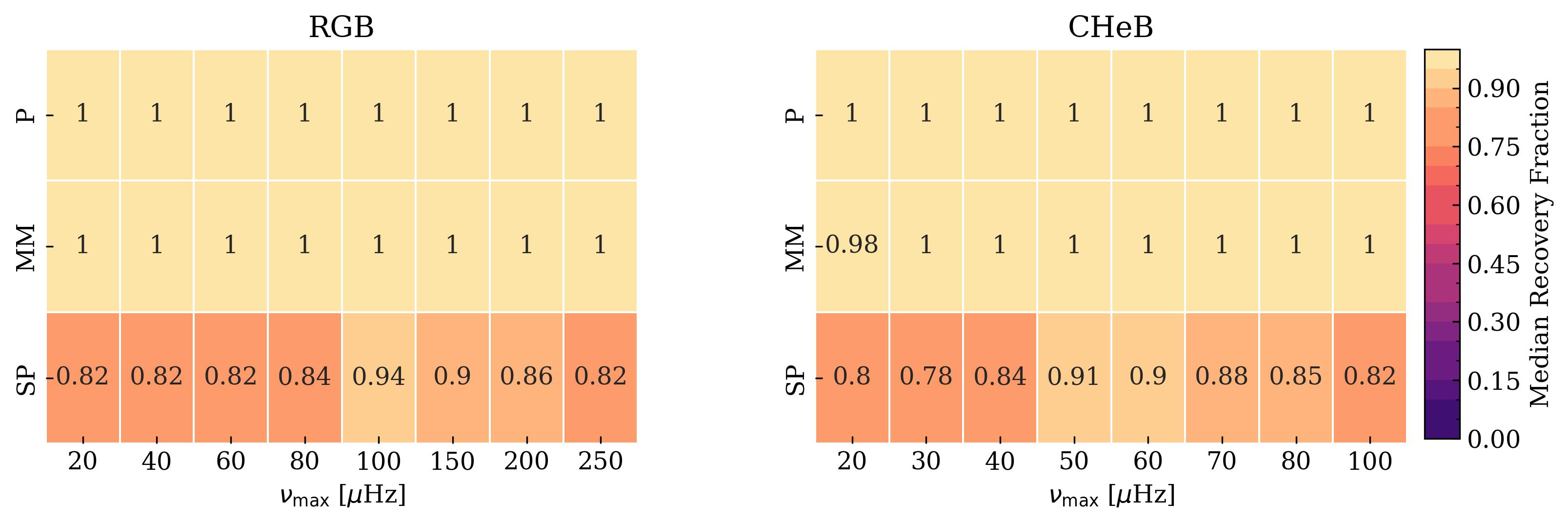}
         \caption{Validation results for synthetic stars with singlets. The left and right panels correspond to RGB and CHeB synthetic stars, respectively. Within each panel, the x-axis represents $\nu_{\rm max}$ and the y-axis the three different tests: performance (P), missing modes (MM), and spurious peaks (SP). Each cell is colour-coded by the median recovery fraction, representing the ratio of correctly identified modes to the total number of modes in the synthetic star. 
         }\label{fig:heatmap_singlets}
      \end{figure*}
      
   \subsection{Framework for statistical validation}\label{subsec:framework_validation}
      We evaluated the method through three tests designed to assess the robustness of the pipeline and to mimic observational challenges. For each synthetic star in our grid, we performed 100 independent realisations of each of these tests:

      \paragraph{GA performance (P)} First, we tested the inherent stability of the GA. For each synthetic star, we performed 100 optimisation runs on the same frequency set to ensure that the stochastic nature of the GA consistently converges to the global maximum.

      \paragraph{Missing modes (MM)} To simulate incomplete mode detection, we implemented a probabilistic removal scheme. For each synthetic star, we first randomly selected acoustic orders to be affected (with a 50\% probability per order). Within an affected order, we determined the number of modes to remove using a binomial distribution. To mimic the observational bias where modes with higher mode inertia (more $g$-dominated) are harder to detect, we weighted the removal probability by the frequency distance from the nominal $p$ mode. This approach led to cases with up to 65\% of modes removed from the frequency set.

      \paragraph{Spurious peaks (SP)} We tested the method's robustness against background noise by introducing spurious frequencies. To evaluate how noise can systematically jeopardise the correct identification, we varied the number of spurious peaks added to each synthetic star from one to ten (i.e. running 10 consecutive iterations for each integer number of added peaks from 1 to 10). We drew the frequencies for these spurious peaks from a uniform distribution spanning the star's oscillation frequency range.\\
   
      In total, this validation represents 12,800 optimisation runs for each test, totalling 38,400 runs across the entire grid (see Figs.~\ref{fig:heatmap_singlets}, \ref{fig:heatmap_doublets}, and \ref{fig:heatmap_triplets}). To quantify the performance, we defined the ``recovery fraction'' per synthetic star as the ratio of correctly identified $\ell=1$ modes to the total number of $\ell=1$ modes in the synthetic star. A mode was considered correctly identified if its frequency matched the true value within a defined tolerance threshold (i.e. $3\sigma_{\rm max}$), and its azimuthal order and $n$ were correctly assigned.
      
   \subsection{Results}
\begin{figure*}
            \centering
            \includegraphics[width=1.0\textwidth]{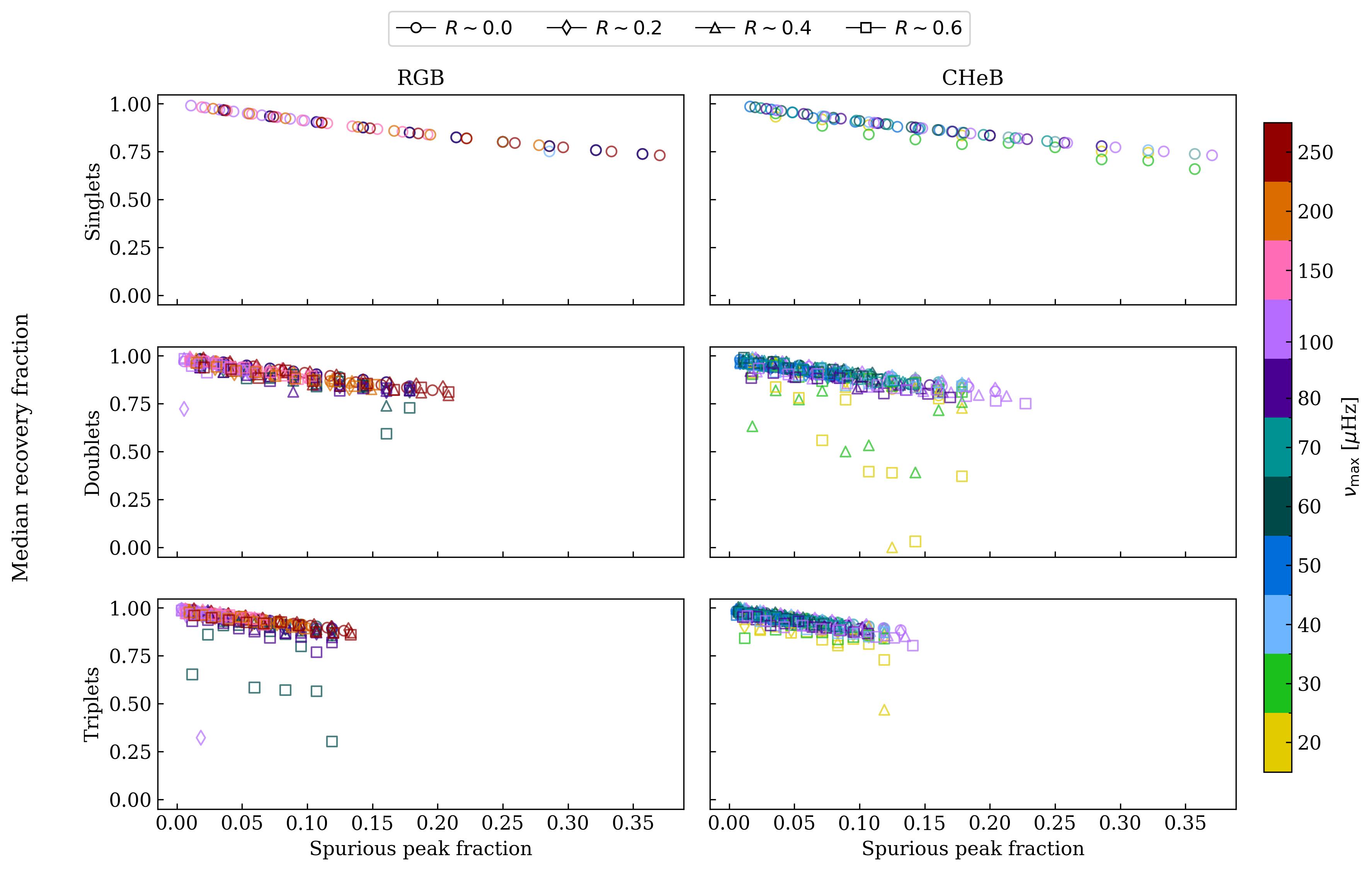}
            \caption{Median recovery fraction as a function of the fraction of spurious peaks for synthetic RGB (left) and CHeB (right) stars. The rows show results for cases with singlets (top), doublets (middle), and triplets (bottom). Different symbols indicate the different crossover ratios as shown in Fig.~\ref{fig:rot_regimes}, while the colours indicate $\nu_{\rm max}$.} \label{fig:sp_trend}
      \end{figure*}

      The results of each test are presented in Figs.~\ref{fig:heatmap_singlets}, \ref{fig:heatmap_doublets}, and \ref{fig:heatmap_triplets}. First, the GA performance test demonstrates that our method remains highly robust thanks to the stochastic nature of the optimisation algorithm. Overall, we achieve a typical recovery fraction above 0.95, with the best performance found in stars with undetectable rotational splittings (i.e. singlets). Second, the method is highly robust in the missing modes test, demonstrating a promising capability for real observations, where it is common to detect incomplete sets of frequencies, whether due to low signal-to-noise ratios or the stochastic nature of mode excitation. Finally, the spurious peaks test reveals that adding artificial noise peaks systematically decreases the median recovery fraction, as illustrated in Fig.~\ref{fig:sp_trend}. We also observe a larger scatter in the recovery fraction for stars with doublets and triplets with $R\geq0.4$ compared to singlets. This happens because the higher mode density of doublets and triplets increases the probability of incorrectly matching a noise peak to a theoretical multiplet component. Nevertheless, the algorithm remains robust, with a typical recovery fraction above 0.85. We find no significant difference in the method's performance between RGB and CHeB stars.
   
      Finally, our method reliably identifies dipole mixed modes across the typical rotational regimes observed in both populations. 

\section{Application to \emph{Kepler} red giants}\label{sec:real_stars}
   As a final test, we applied the method to a set of benchmark red-giant stars observed by \emph{Kepler}, which have been extensively analysed in the literature. The first star, KIC~9145955, is a bright RGB star whose spectrum does not present rotational splittings \citep[e.g.][]{Datta15,Zhang18, Zhang22,Gehan24, Wang25}. KIC~12008916 and KIC~4448777 are low-luminosity RGB stars with clear rotational splittings \citep[e.g.][]{Corsaro15b,DiMauro16, DiMauro18, Corsaro20, LiGang24}. Finally, KIC~5000307 is a Li-rich CHeB star where rotational splittings have been reported by \citet{SilvaAguirre14} and \citet{Mosser24}. We defined the kernel widths $\hat{\sigma}$ for each star based on the line-width distribution of its $\ell=1$ candidates, similarly to Sect.~\ref{subsec:sigma}. More specifically, we set the lower bound to the maximum between the frequency resolution of the observations and the 10th percentile of the star's dipole-mode line-width distribution. We set the upper bound to the 95th percentile of these line widths, capped at 0.2~$\mu$Hz.
   
   \begin{figure*}[ht]
      \centering
      \includegraphics[width=0.45\textwidth]{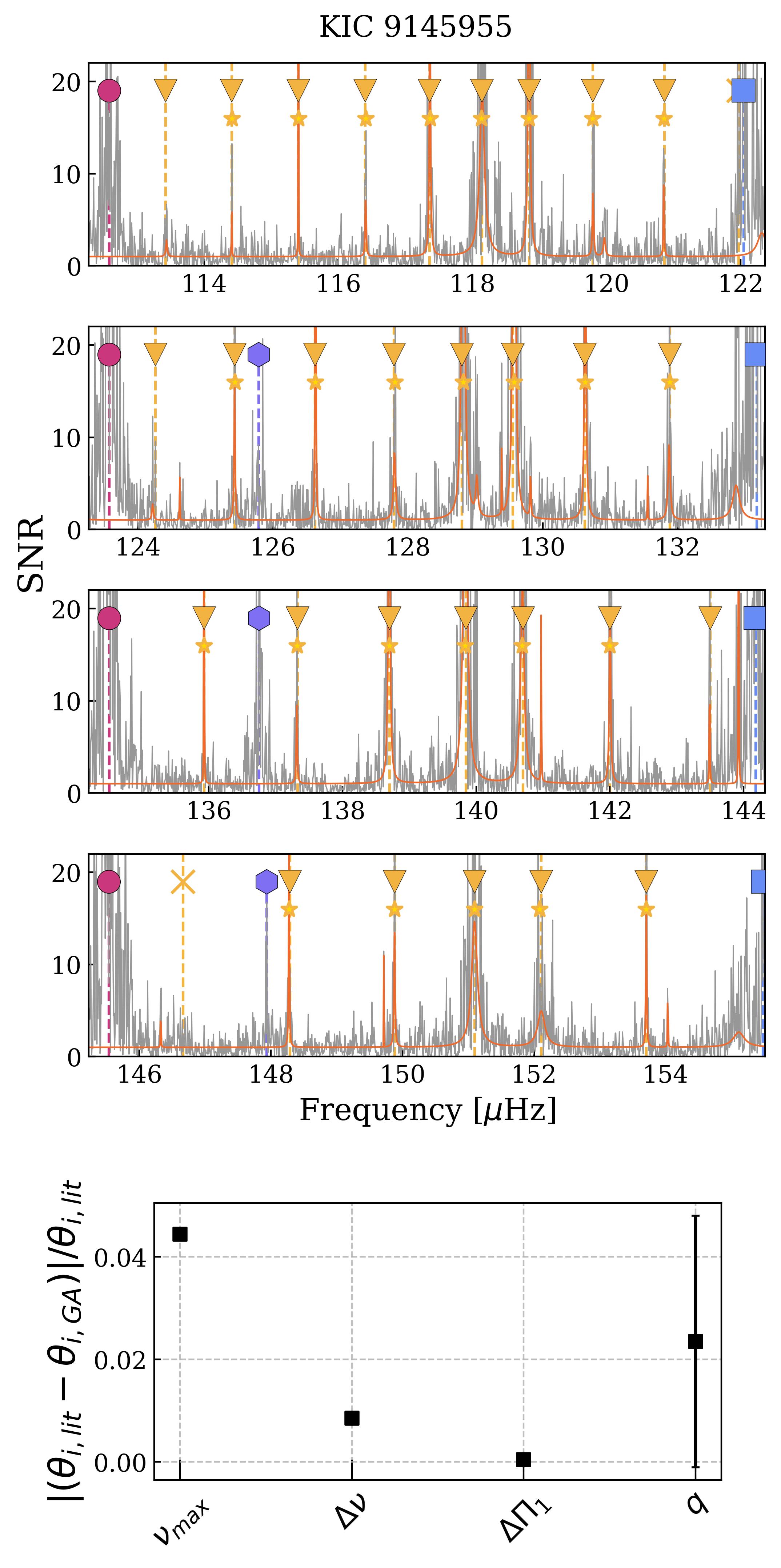}
      \includegraphics[width=0.436\textwidth]{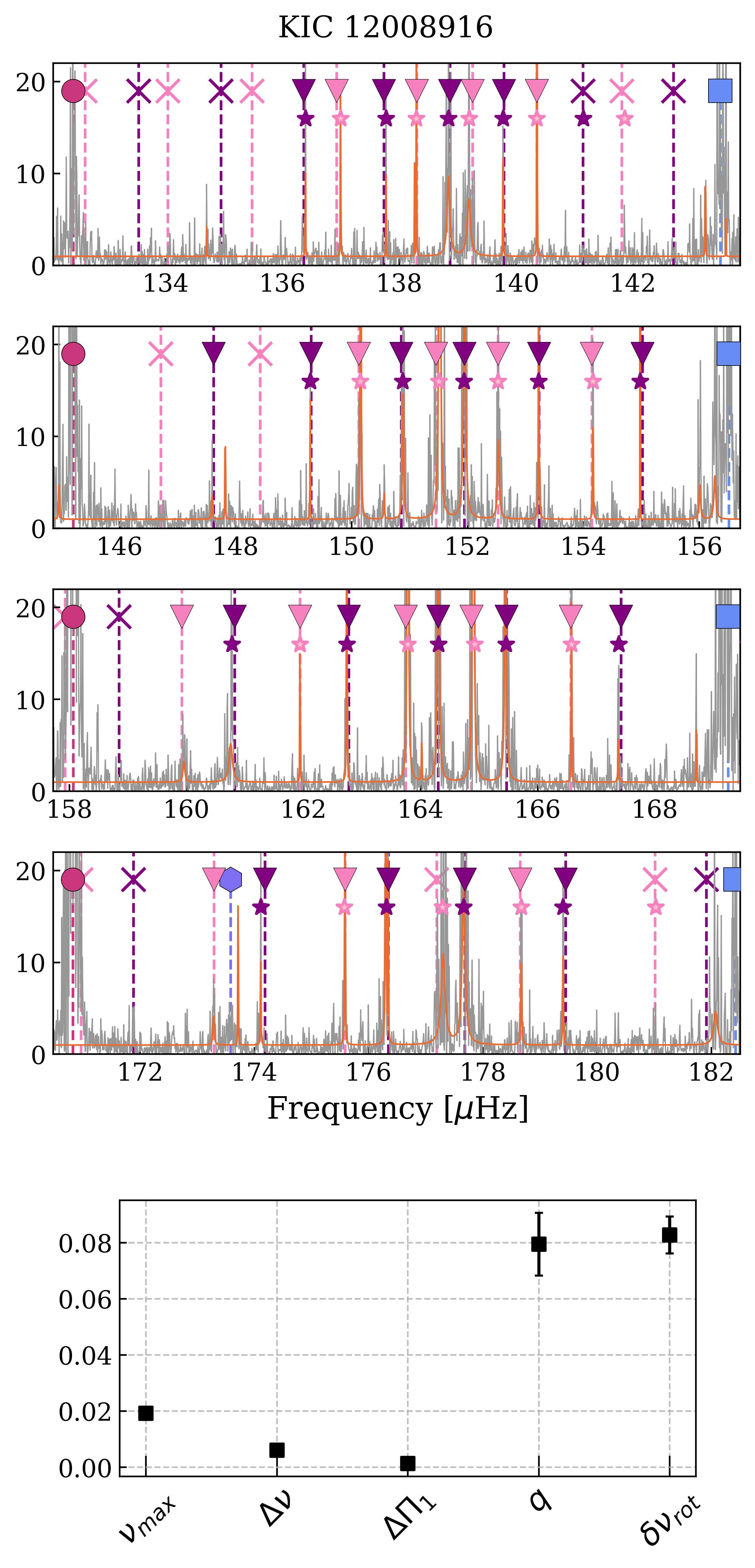}
      \caption{Top: Dipole mode identification for KIC~9145955 (left) and KIC~12008916 (right), showing the four most central acoustic orders. The PDS and the TACO fit to the $\ell=1$ mode candidates are plotted in grey and orange, respectively. Radial, quadrupole, and octupole modes identified by TACO are marked with pink circles, blue squares, and indigo hexagons. Dipole azimuthal orders ($m=-1, 0, 1$) identified in this work are indicated by light pink, yellow, and purple triangles. Crosses show predicted but undetected dipole modes. Stars indicate dipole mixed modes reported in \citet{LiGang24}. Bottom: Relative differences in the parameters of the mixed-mode equation, $\theta_{i}$, between our results and literature values from \citet{LiGang24}, as detailed in Table~\ref{table:comparison_literature}.}\label{fig:kepler_rgs}
      \end{figure*}

      \begin{figure*}[ht]
      \centering
      \includegraphics[width=0.45\textwidth]{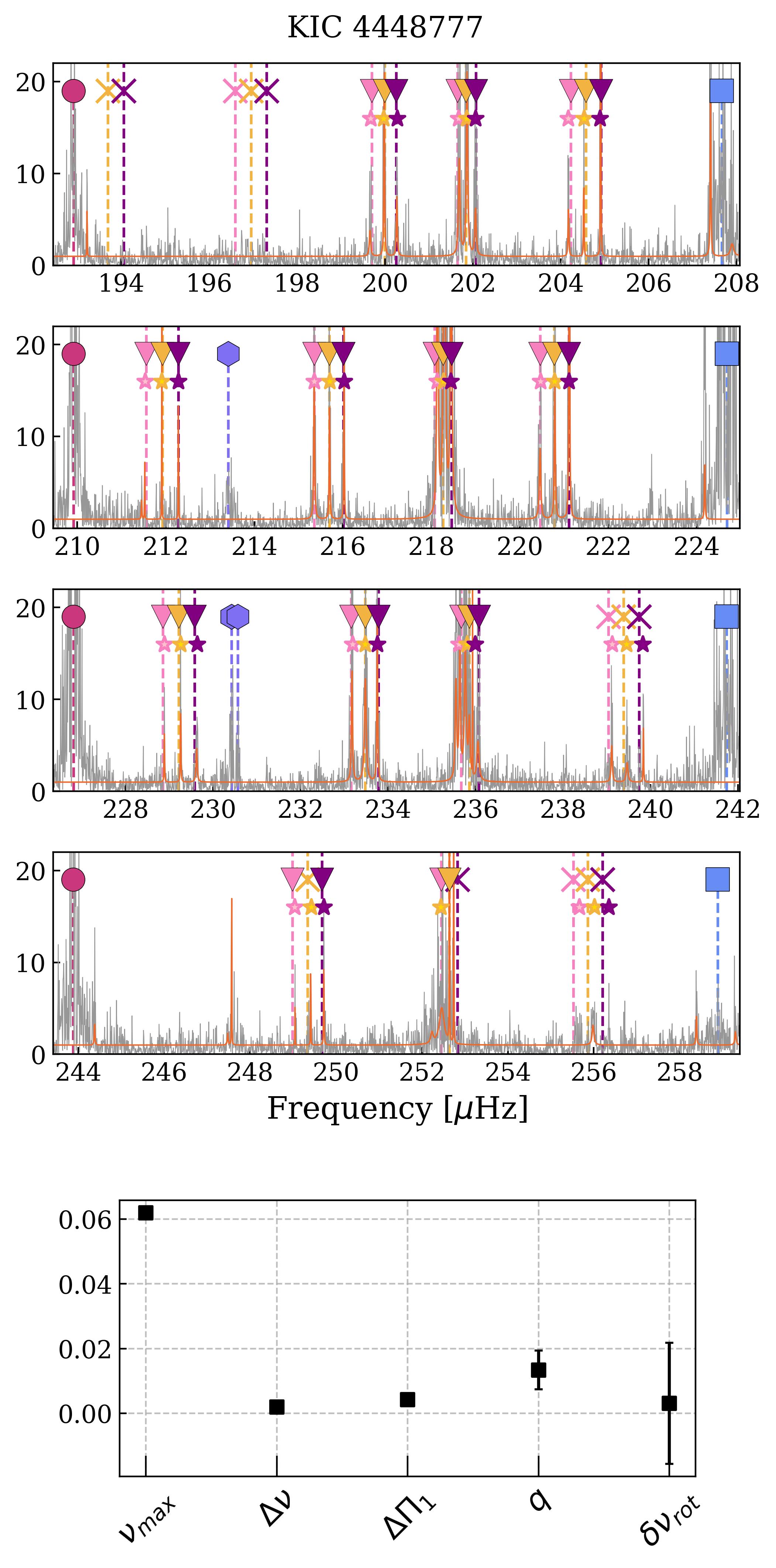}
      \includegraphics[width=0.45\textwidth]{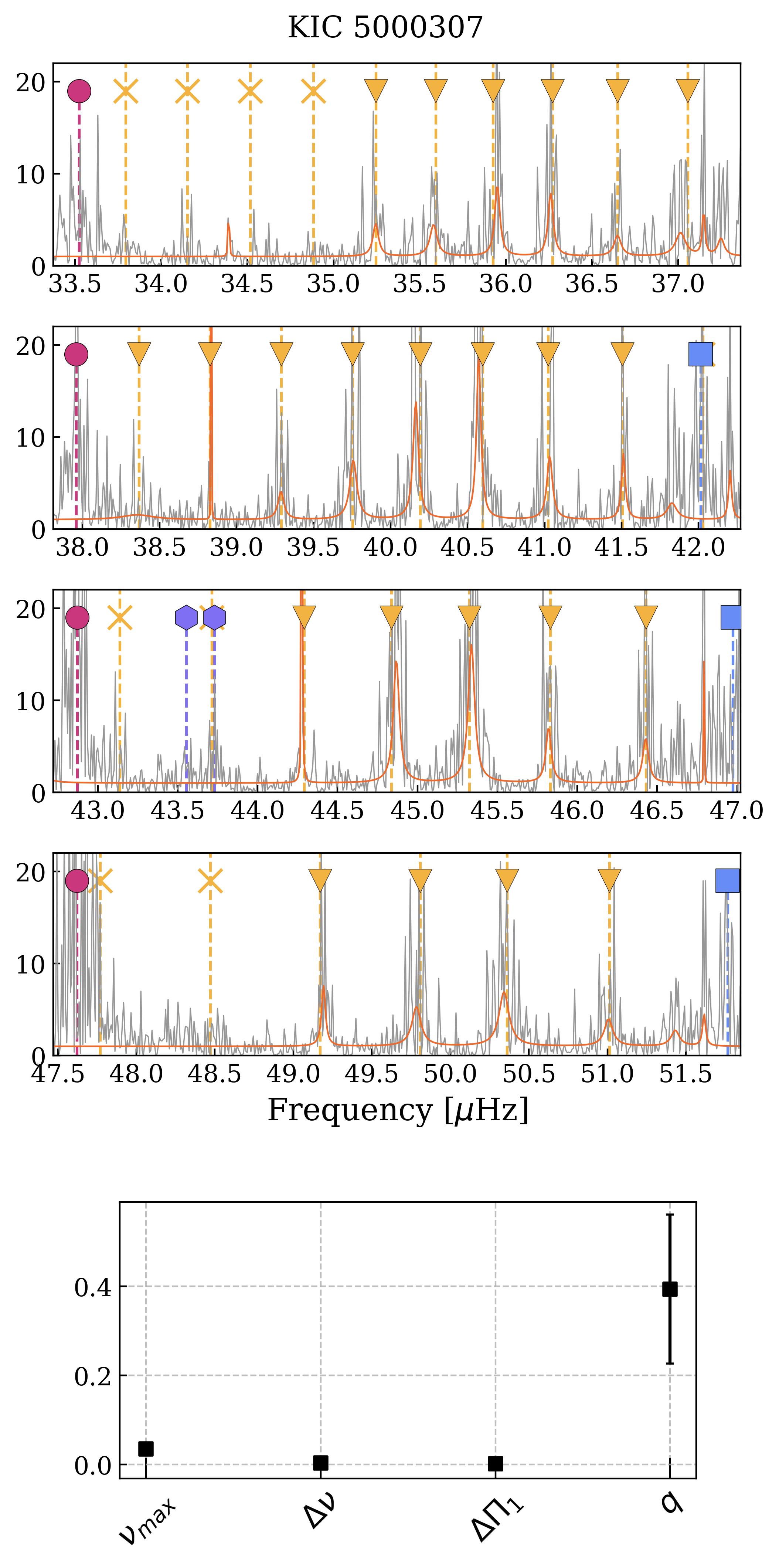}
      \caption{Same as Fig.~\ref{fig:kepler_rgs} but for KIC~4448777 and KIC~5000307. The comparison in the bottom panels is now done with \citet{LiGang24} for KIC~4448777, and with \citet{Mosser24} for KIC~5000307.}\label{fig:kepler_rgs2}
   \end{figure*}

   \subsection{Asteroseismic parameters and oscillation frequencies}
      We derived the global asteroseismic parameters and oscillation-mode properties (frequencies, line widths, and amplitudes) using the TACO pipeline. TACO provides estimates of $\nu_{\rm max}$ and $\Delta\nu$, the phase offset $\epsilon_{p}$, the curvature term $\alpha$, and the identification of $\ell=0, 2, 3$ modes. Peaks not classified as $\ell=0, 2, 3$ are treated as dipole-mode candidates. These candidates were used to obtain an initial estimate of $\Delta \Pi_{1}$ from the power spectrum of the stretched period spectrum. 

      We defined the initial guesses for the parameters of the mixed-mode equation based on the TACO results. If no initial guess for $\Delta \Pi_{1}$ was available from TACO, we adopted an expected value based on the evolutionary state and the scaling relation shown in Appendix~\ref{appendix:scaling_relations}, following the classification described in \citet{Kallinger12}. For $\epsilon_{g}$, $d_{01}$, and the rotational splitting $\delta\nu_{\rm rot}$, we adopted average values inferred from the distributions reported by \citet{LiGang24} and \citet{Gehan18}. Initial guesses for the coupling factor $q$ were set using its linear relation with $\Delta\nu$ for RGB stars (see Appendix~\ref{appendix:scaling_relations}), and a fixed value of $q=0.3$ for CHeB stars, based on the distribution reported by \citet{Vrard16}.

   \subsection{Identified dipole modes}
      Figures~\ref{fig:kepler_rgs} and \ref{fig:kepler_rgs2} illustrate the performance of the algorithm on the four benchmark stars. The top panels show the resulting best-fit asymptotic frequencies compared to the $\ell=1$ mode candidates extracted by TACO. Notably, the presence of spurious noise peaks around the $\ell=1$ candidates (particularly visible in KIC~9145955) does not bias the fit. Furthermore, the pipeline is robust against missing modes. For instance, in KIC~4448777, some dipole modes predicted by the asymptotic relation coincide with visible power excesses in the spectrum (e.g. between 255 and 257~$\mu$Hz). However, because these peaks did not meet the statistical significance threshold during TACO's extraction, they were excluded from the input candidates. Despite this, the method correctly infers the underlying frequency pattern.

      For comparison, we also plotted the points in the power spectrum\footnote{The authors do not report the precise frequencies for these modes.} where \citet{LiGang24} identified dipole mixed modes in KIC~9145955, KIC~4448777, and KIC~12008916. We find excellent agreement in both the frequencies and the azimuthal orders in all cases. Additionally, we compared our results for KIC~4448777 with \citet{DiMauro18} and for KIC~12008916 with \citet{Corsaro15b}. In both studies, the authors used the extended {\sc DIAMONDS} pipeline \citep{Corsaro14,Corsaro15b}, a non-automated approach in which candidate peaks are manually selected, fitted, tested against the noise level, and subsequently identified. For both stars, our automated mode identification is consistent with their results for all common peaks. However, \citet{DiMauro18} reported more dipole modes for KIC~4448777, likely because they manually selected more candidate peaks than TACO automatically detected. To the best of our knowledge, no individual mode frequencies have previously been reported in the literature for KIC~5000307.

      We note that the mixed-mode equation does not account for the effects of sharp structural variations in the stellar structure on the period spacing of pure $g$ modes \citep[e.g.][]{Cunha15,Cunha19,Matteuzzi25,Schimak26}. Hence, we checked the applicability of our method on two clump stars with signatures of buoyancy glitches reported in \citet{Vrard22}. We found that, thanks to the frequency tolerance incorporated into the Gaussian kernel width, our approach can still identify the dipole modes, provided the frequency shifts are smaller than the maximum frequency tolerance (see Appendix~\ref{appendix:glitches}). A more comprehensive analysis of a larger set of such stars will be presented in a forthcoming paper.

   \subsection{Parameters of the mixed-mode equation}
      The parameters of the mixed-mode equation obtained in this work are consistent with those reported in the literature (see Table~\ref{table:comparison_literature} and the bottom panels of Figs.~\ref{fig:kepler_rgs} and \ref{fig:kepler_rgs2}). In particular, the large frequency separation, $\Delta\nu$, agrees with values in the literature to within $1\%$, while $\nu_{\rm max}$ differs by less than $7\%$. This level of agreement is consistent with the typical scatter observed between independent pipelines \citep[e.g.][]{Hekker11a, Hekker12, Pinsonneault18}. The recovered period spacings, $\Delta\Pi_1$, show excellent agreement, with relative differences below $0.5\%$ for all stars. This demonstrates that the method accurately determines the characteristic spacing of the mixed modes. For the stars exhibiting clear rotational splittings (KIC~12008916 and KIC~4448777), the recovered $\delta\nu_{\rm rot}$ agrees with literature estimates to within $10\%$. Larger relative differences are found for phase-related parameters, such as $\epsilon_{p}$ and $\epsilon_{g}$. This behaviour is expected, since $\tilde{\mathcal{S}}_{\rm mbw}$ is inherently less sensitive to variations in phase offsets, as demonstrated in Sect.~\ref{subsec:score_sensitivity}. 

      The inferred coupling factors, $q$, are consistent with published values for the RGB stars (with relative differences typically $\lesssim 3\%$), though a larger discrepancy of around $39\%$ is observed for the CHeB star KIC~5000307 with respect to the results reported by \citet{Mosser24}. However, after visual inspection of Fig.~\ref{fig:kepler_rgs2} (right panel), we confirm that our best fit successfully captures the observed peaks. Therefore, this disagreement with \citet{Mosser24} likely arises from the low sensitivity to $q$ of the stretched-period échelle method used in their work, as noted by \citet{Mosser15} and \citet{Ong23}.
      
      Interestingly, \citet{SilvaAguirre14} and \citet{Mosser24} reported evidence of stellar rotation in KIC~5000307, estimating $\delta\nu_{\rm rot} \approx 50~$nHz and $\delta\nu_{\rm rot} \approx 42\pm10~$nHz, respectively. Since our method is designed to operate as a higher-level module of TACO, its sensitivity is inherently tied to the initial peak-bagging performance. In this case, TACO detected only a single candidate frequency per mixed mode, and therefore the algorithm lacked the necessary inputs to fit a rotational splitting. Nevertheless, the modular nature of our method ensures it can easily be adapted to work with inputs from other peak-extraction tools, provided they supply the frequencies of the dipole mode candidates and radial modes.

\section{Conclusions}\label{sec:discussion}
   In this work, we have presented a robust and automated method for identifying dipole mixed modes in red-giant stars. This method is based on the multi-bandwidth score ($\tilde{\mathcal{S}}_{\rm mbw}$), a metric that measures the overlap between theoretical and observed convolved spike trains. We showed that by optimising this metric using a GA, the method addresses common challenges in mixed-mode identification. These include the presence of noise peaks, incomplete mode detection, and degeneracies inherent in the asymptotic parameter space, further enhanced by rotational splittings. 

   We evaluated the method's performance using a framework based on \emph{Kepler} data, with the Gaussian kernel widths calibrated using empirical bounds derived from the CAPASS catalogue (Espinoza-Rojas et al., in prep.). The validation across 38,400 synthetic stars in different evolutionary states, showed a typical median recovery fraction (the fraction of correctly identified dipole modes) above 0.95. Although we find the performance to scale inversely with the fraction of spurious peaks, the median recovery fraction remains above 0.85 in the presence of noise. Regarding computational complexity, the GA typically converges within 30 minutes to a few hours per star on a standard laptop.
   
   Moreover, the test on benchmark \emph{Kepler} targets confirms that our approach is well suited for the analysis of real observational data. We find that the dipole modes recovered by our method are consistent with previously published results, and that our estimates for the parameters of the mixed-mode equation typically agree with the literature within $1\%$ for $\Delta\Pi_{1}$, and within $10\%$ for $\delta\nu_{\rm rot}$ and $q$. Furthermore, our method is flexible enough to account for structural glitches, provided the deviations fall within the tolerance of the kernel width. Further analysis of such cases will be presented in a separate publication.

   The multi-bandwidth score, $\tilde{\mathcal{S}}_{\rm mbw}$, is designed to measure how accurately the observations are described by the mixed-mode equation, rather than to serve as a formal likelihood. This design choice allows the GA to efficiently locate a robust global solution in a complex parameter space. Consequently, the best-fit parameters provide well-constrained initial guesses for subsequent optimisation routines to estimate formal uncertainties. Since the main goal of this method is to identify dipole modes rather than to provide precise estimates of the asymptotic parameters, we leave this implementation for future work. Additionally, while our treatment of rotational asymmetries is empirical, this approach ensures that we capture the observed frequency patterns without imposing strict theoretical assumptions. Accurately extracting these patterns is an essential first step for investigating the underlying physical mechanisms that drive such deviations.
   
   Since the method presented in this work is completely data-driven and robust against missing modes and spurious peaks, it is well suited for the automated and homogeneous seismic characterisation of evolved stars. When integrated as a module within extraction pipelines such as TACO, it provides a robust framework for processing the extensive datasets expected from current missions and upcoming large-scale surveys such as PLATO \citep{Rauer25} or the Earth 2.0 Space Mission \citep{Ge22}.

   \begin{acknowledgements}
   We would like to thank the anonymous referee for their insightful comments and suggestions. We acknowledge funding from the ERC Consolidator Grant DipolarSound (grant agreement \#101000296). We also thank the Klaus Tschira Foundation for their support. This paper includes data collected by the \emph{Kepler} mission and obtained from the MAST data archive at the Space Telescope Science Institute (STScI). Funding for the \emph{Kepler} mission is provided by the NASA Science Mission Directorate. STScI is operated by the Association of Universities for Research in Astronomy, Inc., under NASA contract NAS 5–26555. F.E.R. thanks Felix Ahlborn for his important insights and comments on the manuscript, and Jordan Van Beeck and Jeong Yun Choi for the fruitful discussions.

\end{acknowledgements}

\bibliographystyle{aa}
\bibliography{bibliography}

\begin{thebibliography}{100}
\expandafter\ifx\csname natexlab\endcsname\relax\def\natexlab#1{#1}\fi

\bibitem[{{Aerts} {et~al.}(2010){Aerts}, {Christensen-Dalsgaard}, \& {Kurtz}}]{Aerts10}
{Aerts}, C., {Christensen-Dalsgaard}, J., \& {Kurtz}, D.~W. 2010, {Asteroseismology}

\bibitem[{{Ahlborn} {et~al.}(2020){Ahlborn}, {Bellinger}, {Hekker}, {Basu}, \& {Angelou}}]{Ahlborn20}
{Ahlborn}, F., {Bellinger}, E.~P., {Hekker}, S., {Basu}, S., \& {Angelou}, G.~C. 2020, \aap, 639, A98

\bibitem[{{Ahlborn} {et~al.}(2025){Ahlborn}, {Joel Ong}, {Van Beeck}, {Bellinger}, {Hekker}, \& {Basu}}]{Ahlborn25}
{Ahlborn}, F., {Joel Ong}, J.~M., {Van Beeck}, J., {et~al.} 2025, \aap, 704, A230

\bibitem[{{Baglin} {et~al.}(2006){Baglin}, {Auvergne}, {Barge}, {Deleuil}, {Catala}, {Michel}, {Weiss}, \& {COROT Team}}]{CoRoT}
{Baglin}, A., {Auvergne}, M., {Barge}, P., {et~al.} 2006, in ESA Special Publication, Vol. 1306, The CoRoT Mission Pre-Launch Status - Stellar Seismology and Planet Finding, ed. M.~{Fridlund}, A.~{Baglin}, J.~{Lochard}, \& L.~{Conroy}, 33

\bibitem[{{Beck} {et~al.}(2012){Beck}, {Montalban}, {Kallinger}, {De Ridder}, {Aerts}, {Garc{\'\i}a}, {Hekker}, {Dupret}, {Mosser}, {Eggenberger}, {Stello}, {Elsworth}, {Frandsen}, {Carrier}, {Hillen}, {Gruberbauer}, {Christensen-Dalsgaard}, {Miglio}, {Valentini}, {Bedding}, {Kjeldsen}, {Girouard}, {Hall}, \& {Ibrahim}}]{Beck12}
{Beck}, P.~G., {Montalban}, J., {Kallinger}, T., {et~al.} 2012, \nat, 481, 55

\bibitem[{{Bedding} {et~al.}(2011){Bedding}, {Mosser}, {Huber}, {Montalb{\'a}n}, {Beck}, {Christensen-Dalsgaard}, {Elsworth}, {Garc{\'\i}a}, {Miglio}, {Stello}, {White}, {De Ridder}, {Hekker}, {Aerts}, {Barban}, {Belkacem}, {Broomhall}, {Brown}, {Buzasi}, {Carrier}, {Chaplin}, {di Mauro}, {Dupret}, {Frandsen}, {Gilliland}, {Goupil}, {Jenkins}, {Kallinger}, {Kawaler}, {Kjeldsen}, {Mathur}, {Noels}, {Silva Aguirre}, \& {Ventura}}]{Bedding11}
{Bedding}, T.~R., {Mosser}, B., {Huber}, D., {et~al.} 2011, \nat, 471, 608

\bibitem[{{Bordad{\'a}gua} {et~al.}(2025){Bordad{\'a}gua}, {Ahlborn}, {Copp{\'e}e}, {Marques}, {Belkacem}, \& {Hekker}}]{Bordadagua25}
{Bordad{\'a}gua}, B., {Ahlborn}, F., {Copp{\'e}e}, Q., {et~al.} 2025, \aap, 699, A310

\bibitem[{{Borucki} {et~al.}(2010){Borucki}, {Koch}, {Basri}, {Batalha}, {Brown}, {Caldwell}, {Caldwell}, {Christensen-Dalsgaard}, {Cochran}, {DeVore}, {Dunham}, {Dupree}, {Gautier}, {Geary}, {Gilliland}, {Gould}, {Howell}, {Jenkins}, {Kondo}, {Latham}, {Marcy}, {Meibom}, {Kjeldsen}, {Lissauer}, {Monet}, {Morrison}, {Sasselov}, {Tarter}, {Boss}, {Brownlee}, {Owen}, {Buzasi}, {Charbonneau}, {Doyle}, {Fortney}, {Ford}, {Holman}, {Seager}, {Steffen}, {Welsh}, {Rowe}, {Anderson}, {Buchhave}, {Ciardi}, {Walkowicz}, {Sherry}, {Horch}, {Isaacson}, {Everett}, {Fischer}, {Torres}, {Johnson}, {Endl}, {MacQueen}, {Bryson}, {Dotson}, {Haas}, {Kolodziejczak}, {Van Cleve}, {Chandrasekaran}, {Twicken}, {Quintana}, {Clarke}, {Allen}, {Li}, {Wu}, {Tenenbaum}, {Verner}, {Bruhweiler}, {Barnes}, \& {Prsa}}]{Borucki10}
{Borucki}, W.~J., {Koch}, D., {Basri}, G., {et~al.} 2010, Science, 327, 977

\bibitem[{{Buysschaert} {et~al.}(2016){Buysschaert}, {Beck}, {Corsaro}, {Christensen-Dalsgaard}, {Aerts}, {Arentoft}, {Kjeldsen}, {Garc{\'\i}a}, {Silva Aguirre}, \& {Degroote}}]{Buysschaert16}
{Buysschaert}, B., {Beck}, P.~G., {Corsaro}, E., {et~al.} 2016, \aap, 588, A82

\bibitem[{{Cantiello} {et~al.}(2014){Cantiello}, {Mankovich}, {Bildsten}, {Christensen-Dalsgaard}, \& {Paxton}}]{Cantiello14}
{Cantiello}, M., {Mankovich}, C., {Bildsten}, L., {Christensen-Dalsgaard}, J., \& {Paxton}, B. 2014, \apj, 788, 93

\bibitem[{{Choi} {et~al.}(2025){Choi}, {Espinoza-Rojas}, {Copp{\'e}e}, \& {Hekker}}]{Choi25}
{Choi}, J.~Y., {Espinoza-Rojas}, F., {Copp{\'e}e}, Q., \& {Hekker}, S. 2025, \aap, 699, A180

\bibitem[{{Christensen-Dalsgaard} {et~al.}(2014){Christensen-Dalsgaard}, {Silva Aguirre}, {Elsworth}, \& {Hekker}}]{JCD14}
{Christensen-Dalsgaard}, J., {Silva Aguirre}, V., {Elsworth}, Y., \& {Hekker}, S. 2014, \mnras, 445, 3685

\bibitem[{{Corsaro} \& {De Ridder}(2014)}]{Corsaro14}
{Corsaro}, E. \& {De Ridder}, J. 2014, \aap, 571, A71

\bibitem[{{Corsaro} {et~al.}(2015b){Corsaro}, {De Ridder}, \& {Garc{\'\i}a}}]{Corsaro15b}
{Corsaro}, E., {De Ridder}, J., \& {Garc{\'\i}a}, R.~A. 2015b, \aap, 579, A83

\bibitem[{{Corsaro} {et~al.}(2017){Corsaro}, {Lee}, {Garc{\'\i}a}, {Hennebelle}, {Mathur}, {Beck}, {Mathis}, {Stello}, \& {Bouvier}}]{Corsaro17}
{Corsaro}, E., {Lee}, Y.-N., {Garc{\'\i}a}, R.~A., {et~al.} 2017, Nature Astronomy, 1, 0064

\bibitem[{{Corsaro} {et~al.}(2020){Corsaro}, {McKeever}, \& {Kuszlewicz}}]{Corsaro20}
{Corsaro}, E., {McKeever}, J.~M., \& {Kuszlewicz}, J.~S. 2020, \aap, 640, A130

\bibitem[{{Cunha} {et~al.}(2019){Cunha}, {Avelino}, {Christensen-Dalsgaard}, {Stello}, {Vrard}, {Jiang}, \& {Mosser}}]{Cunha19}
{Cunha}, M.~S., {Avelino}, P.~P., {Christensen-Dalsgaard}, J., {et~al.} 2019, \mnras, 490, 909

\bibitem[{{Cunha} {et~al.}(2015){Cunha}, {Stello}, {Avelino}, {Christensen-Dalsgaard}, \& {Townsend}}]{Cunha15}
{Cunha}, M.~S., {Stello}, D., {Avelino}, P.~P., {Christensen-Dalsgaard}, J., \& {Townsend}, R.~H.~D. 2015, \apj, 805, 127

\bibitem[{{Datta} {et~al.}(2015){Datta}, {Mazumdar}, {Gupta}, \& {Hekker}}]{Datta15}
{Datta}, A., {Mazumdar}, A., {Gupta}, U., \& {Hekker}, S. 2015, \mnras, 447, 1935

\bibitem[{{Deheuvels} {et~al.}(2015){Deheuvels}, {Ballot}, {Beck}, {Mosser}, {{\O}stensen}, {Garc{\'\i}a}, \& {Goupil}}]{Deheuvels15}
{Deheuvels}, S., {Ballot}, J., {Beck}, P.~G., {et~al.} 2015, \aap, 580, A96

\bibitem[{{Deheuvels} {et~al.}(2014){Deheuvels}, {Do{\u{g}}an}, {Goupil}, {Appourchaux}, {Benomar}, {Bruntt}, {Campante}, {Casagrande}, {Ceillier}, {Davies}, {De Cat}, {Fu}, {Garc{\'\i}a}, {Lobel}, {Mosser}, {Reese}, {Regulo}, {Schou}, {Stahn}, {Thygesen}, {Yang}, {Chaplin}, {Christensen-Dalsgaard}, {Eggenberger}, {Gizon}, {Mathis}, {Molenda-{\.Z}akowicz}, \& {Pinsonneault}}]{Deheuvels14}
{Deheuvels}, S., {Do{\u{g}}an}, G., {Goupil}, M.~J., {et~al.} 2014, \aap, 564, A27

\bibitem[{{Deheuvels} {et~al.}(2012){Deheuvels}, {Garc{\'\i}a}, {Chaplin}, {Basu}, {Antia}, {Appourchaux}, {Benomar}, {Davies}, {Elsworth}, {Gizon}, {Goupil}, {Reese}, {Regulo}, {Schou}, {Stahn}, {Casagrande}, {Christensen-Dalsgaard}, {Fischer}, {Hekker}, {Kjeldsen}, {Mathur}, {Mosser}, {Pinsonneault}, {Valenti}, {Christiansen}, {Kinemuchi}, \& {Mullally}}]{Deheuvels12}
{Deheuvels}, S., {Garc{\'\i}a}, R.~A., {Chaplin}, W.~J., {et~al.} 2012, \apj, 756, 19

\bibitem[{{Deheuvels} {et~al.}(2017){Deheuvels}, {Ouazzani}, \& {Basu}}]{Deheuvels17}
{Deheuvels}, S., {Ouazzani}, R.~M., \& {Basu}, S. 2017, \aap, 605, A75

\bibitem[{{Di Mauro} {et~al.}(2016){Di Mauro}, {Ventura}, {Cardini}, {Stello}, {Christensen-Dalsgaard}, {Dziembowski}, {Patern{\`o}}, {Beck}, {Bloemen}, {Davies}, {De Smedt}, {Elsworth}, {Garc{\'\i}a}, {Hekker}, {Mosser}, \& {Tkachenko}}]{DiMauro16}
{Di Mauro}, M.~P., {Ventura}, R., {Cardini}, D., {et~al.} 2016, \apj, 817, 65

\bibitem[{{Di Mauro} {et~al.}(2018){Di Mauro}, {Ventura}, {Corsaro}, \& {Lustosa De Moura}}]{DiMauro18}
{Di Mauro}, M.~P., {Ventura}, R., {Corsaro}, E., \& {Lustosa De Moura}, B. 2018, \apj, 862, 9

\bibitem[{{Dziembowski} \& {Goode}(1992)}]{Dziembowski92}
{Dziembowski}, W.~A. \& {Goode}, P.~R. 1992, \apj, 394, 670

\bibitem[{{Eggenberger} {et~al.}(2012){Eggenberger}, {Montalb{\'a}n}, \& {Miglio}}]{Eggenberger12}
{Eggenberger}, P., {Montalb{\'a}n}, J., \& {Miglio}, A. 2012, \aap, 544, L4

\bibitem[{{Eggenberger} {et~al.}(2022){Eggenberger}, {Moyano}, \& {den Hartogh}}]{Eggenberger22}
{Eggenberger}, P., {Moyano}, F.~D., \& {den Hartogh}, J.~W. 2022, \aap, 664, L16

\bibitem[{{Espinoza-Rojas} {et~al.}(2025){Espinoza-Rojas}, {Theme{\ss}l}, \& {Hekker}}]{FER25}
{Espinoza-Rojas}, F., {Theme{\ss}l}, N., \& {Hekker}, S. 2025, \aap, 703, A66

\bibitem[{{Fuller} {et~al.}(2015){Fuller}, {Cantiello}, {Stello}, {Garcia}, \& {Bildsten}}]{Fuller15}
{Fuller}, J., {Cantiello}, M., {Stello}, D., {Garcia}, R.~A., \& {Bildsten}, L. 2015, Science, 350, 423

\bibitem[{Gad(2023)}]{pygad}
Gad, A.~F. 2023, Multimedia Tools and Applications, 1

\bibitem[{{Garc{\'\i}a} \& {Ballot}(2019)}]{Garcia19}
{Garc{\'\i}a}, R.~A. \& {Ballot}, J. 2019, Living Reviews in Solar Physics, 16, 4

\bibitem[{{Garc{\'\i}a Saravia Ortiz de Montellano} {et~al.}(2018){Garc{\'\i}a Saravia Ortiz de Montellano}, {Hekker}, \& {Theme{\ss}l}}]{GSOM18}
{Garc{\'\i}a Saravia Ortiz de Montellano}, A., {Hekker}, S., \& {Theme{\ss}l}, N. 2018, \mnras, 476, 1470

\bibitem[{{Ge} {et~al.}(2022){Ge}, {Zhang}, {Zang}, {Deng}, {Mao}, {Xie}, {Liu}, {Zhou}, {Willis}, {Huang}, {Howell}, {Feng}, {Zhu}, {Yao}, {Liu}, {Aizawa}, {Zhu}, {Li}, {Ma}, {Ye}, {Yu}, {Xiang}, {Yu}, {Liu}, {Yang}, {Wang}, {Shi}, {Fang}, {Zong}, {Liu}, {Zhang}, {Zhang}, {El-Badry}, {Shen}, {Tam}, {Hu}, {Yang}, {Zou}, {Wu}, {Lei}, {Wei}, {Wu}, {Sun}, {Wang}, {Zhang}, {Xu}, {Yang}, {Li}, {Xiang}, {Wang}, {Wang}, {Zhang}, {Jia}, {Yuan}, {Zhang}, {Xuesong Wang}, {Gan}, {Wang}, {Zhao}, {Liu}, {Wei}, {Kang}, {Yang}, {Qi}, {Liu}, {Zhang}, {Zhu}, {Zhou}, {Zhang}, {Yu}, {Zhang}, {Li}, {Tang}, {Wang}, {Wang}, {Li}, {Cheng}, {Shen}, {Li}, {Pan}, {Yang}, {Gao}, {Song}, {Wang}, {Zhang}, {Chen}, {Wang}, {Zhang}, {Wang}, {Zeng}, {Zheng}, {Zhu}, {Guo}, {Zhang}, {Li}, {Wen}, {Feng}, {Chen}, {Chen}, {Han}, {Yang}, {Wang}, {Duan}, {Huang}, {Liang}, {Bi}, {Gai}, {Ge}, {Guo}, {Huang}, {Li}, {Li}, {Li}, {Yuxi}, {Lu}, {Rix}, {Shi}, {Song}, {Tang}, {Ting}, {Wu}, {Wu}, {Yang}, {Yin}, {Gould}, {Lee}, {Dong}, {Yee}, {Shvartzvald}, {Yang}, {Kuang}, {Zhang}, {Liao}, {Qi}, {Yang}, {Zhang}, {Jiang}, {Ou}, {Li}, {Beck}, {Bedding}, {Campante}, {Chaplin}, {Christensen-Dalsgaard}, {Garc{\'\i}a}, {Gaulme}, {Gizon}, {Hekker}, {Huber}, {Khanna}, {Li}, {Mathur}, {Miglio}, {Mosser}, {Ong}, {Santos}, {Stello}, {Bowman}, {Lares-Martiz}, {Murphy}, {Niu}, {Ma}, {Moln{\'a}r}, {Fu}, {De Cat}, {Su}, \& {consortium}}]{Ge22}
{Ge}, J., {Zhang}, H., {Zang}, W., {et~al.} 2022, arXiv e-prints, arXiv:2206.06693

\bibitem[{{Gehan} {et~al.}(2024){Gehan}, {Godoy-Rivera}, \& {Gaulme}}]{Gehan24}
{Gehan}, C., {Godoy-Rivera}, D., \& {Gaulme}, P. 2024, \aap, 686, A93

\bibitem[{{Gehan} {et~al.}(2021){Gehan}, {Mosser}, {Michel}, \& {Cunha}}]{Gehan21}
{Gehan}, C., {Mosser}, B., {Michel}, E., \& {Cunha}, M.~S. 2021, \aap, 645, A124

\bibitem[{{Gehan} {et~al.}(2018){Gehan}, {Mosser}, {Michel}, {Samadi}, \& {Kallinger}}]{Gehan18}
{Gehan}, C., {Mosser}, B., {Michel}, E., {Samadi}, R., \& {Kallinger}, T. 2018, \aap, 616, A24

\bibitem[{{Gizon} \& {Solanki}(2003)}]{Gizon03}
{Gizon}, L. \& {Solanki}, S.~K. 2003, \apj, 589, 1009

\bibitem[{{Gough}(1990)}]{Gough90}
{Gough}, D.~O. 1990, in Progress of Seismology of the Sun and Stars, ed. Y.~{Osaki} \& H.~{Shibahashi}, Vol. 367, 283

\bibitem[{{Hatt} {et~al.}(2024){Hatt}, {Ong}, {Nielsen}, {Chaplin}, {Davies}, {Deheuvels}, {Ballot}, {Li}, \& {Bugnet}}]{Hatt24}
{Hatt}, E.~J., {Ong}, J.~M.~J., {Nielsen}, M.~B., {et~al.} 2024, \mnras, 534, 1060

\bibitem[{{Hekker} {et~al.}(2011{\natexlab{a}}){Hekker}, {Basu}, {Stello}, {Kallinger}, {Grundahl}, {Mathur}, {Garc{\'\i}a}, {Mosser}, {Huber}, {Bedding}, {Szab{\'o}}, {De Ridder}, {Chaplin}, {Elsworth}, {Hale}, {Christensen-Dalsgaard}, {Gilliland}, {Still}, {McCauliff}, \& {Quintana}}]{Hekker11}
{Hekker}, S., {Basu}, S., {Stello}, D., {et~al.} 2011{\natexlab{a}}, \aap, 530, A100

\bibitem[{{Hekker} {et~al.}(2026){Hekker}, {Elsworth}, {Basu}, {Ahlborn}, {Ball}, {Bellinger}, {Buchele}, \& {Espinoza-Rojas}}]{Hekker26}
{Hekker}, S., {Elsworth}, Y., {Basu}, S., {et~al.} 2026, \aap, 709, A84

\bibitem[{{Hekker} {et~al.}(2011{\natexlab{b}}){Hekker}, {Elsworth}, {De Ridder}, {Mosser}, {Garc{\'\i}a}, {Kallinger}, {Mathur}, {Huber}, {Buzasi}, {Preston}, {Hale}, {Ballot}, {Chaplin}, {R{\'e}gulo}, {Bedding}, {Stello}, {Borucki}, {Koch}, {Jenkins}, {Allen}, {Gilliland}, {Kjeldsen}, \& {Christensen-Dalsgaard}}]{Hekker11a}
{Hekker}, S., {Elsworth}, Y., {De Ridder}, J., {et~al.} 2011{\natexlab{b}}, \aap, 525, A131

\bibitem[{{Hekker} {et~al.}(2012){Hekker}, {Elsworth}, {Mosser}, {Kallinger}, {Chaplin}, {De Ridder}, {Garc{\'\i}a}, {Stello}, {Clarke}, {Hall}, \& {Ibrahim}}]{Hekker12}
{Hekker}, S., {Elsworth}, Y., {Mosser}, B., {et~al.} 2012, \aap, 544, A90

\bibitem[{{Hon} {et~al.}(2019){Hon}, {Stello}, {Garc{\'\i}a}, {Mathur}, {Sharma}, {Colman}, \& {Bugnet}}]{Hon19}
{Hon}, M., {Stello}, D., {Garc{\'\i}a}, R.~A., {et~al.} 2019, \mnras, 485, 5616

\bibitem[{{Howell} {et~al.}(2014){Howell}, {Sobeck}, {Haas}, {Still}, {Barclay}, {Mullally}, {Troeltzsch}, {Aigrain}, {Bryson}, {Caldwell}, {Chaplin}, {Cochran}, {Huber}, {Marcy}, {Miglio}, {Najita}, {Smith}, {Twicken}, \& {Fortney}}]{Howell14}
{Howell}, S.~B., {Sobeck}, C., {Haas}, M., {et~al.} 2014, \pasp, 126, 398

\bibitem[{{Huber} {et~al.}(2010){Huber}, {Bedding}, {Stello}, {Mosser}, {Mathur}, {Kallinger}, {Hekker}, {Elsworth}, {Buzasi}, {De Ridder}, {Gilliland}, {Kjeldsen}, {Chaplin}, {Garc{\'\i}a}, {Hale}, {Preston}, {White}, {Borucki}, {Christensen-Dalsgaard}, {Clarke}, {Jenkins}, \& {Koch}}]{Huber10}
{Huber}, D., {Bedding}, T.~R., {Stello}, D., {et~al.} 2010, \apj, 723, 1607

\bibitem[{{Kallinger}(2019)}]{Kallinger19}
{Kallinger}, T. 2019, arXiv e-prints, arXiv:1906.09428

\bibitem[{{Kallinger} {et~al.}(2012){Kallinger}, {Hekker}, {Mosser}, {De Ridder}, {Bedding}, {Elsworth}, {Gruberbauer}, {Guenther}, {Stello}, {Basu}, {Garc{\'\i}a}, {Chaplin}, {Mullally}, {Still}, \& {Thompson}}]{Kallinger12}
{Kallinger}, T., {Hekker}, S., {Mosser}, B., {et~al.} 2012, \aap, 541, A51

\bibitem[{{Kamiaka} {et~al.}(2018){Kamiaka}, {Benomar}, \& {Suto}}]{Kamiaka18}
{Kamiaka}, S., {Benomar}, O., \& {Suto}, Y. 2018, \mnras, 479, 391

\bibitem[{Kass {et~al.}(2014)Kass, Eden, \& Brown}]{Kass14}
Kass, R.~E., Eden, U.~T., \& Brown, E.~N. 2014, Analysis of Neural Data (New York: Springer)

\bibitem[{Kuhn(1955)}]{Kuhn55}
Kuhn, H.~W. 1955, Naval Research Logistics Quarterly, 2, 83

\bibitem[{Kuszlewicz {et~al.}(2023)Kuszlewicz, Hon, \& Huber}]{Kuszlewicz23}
Kuszlewicz, J.~S., Hon, M., \& Huber, D. 2023, The Astrophysical Journal, 954, 152

\bibitem[{{Li} {et~al.}(2024){Li}, {Deheuvels}, \& {Ballot}}]{LiGang24}
{Li}, G., {Deheuvels}, S., \& {Ballot}, J. 2024, \aap, 688, A184

\bibitem[{{Li} {et~al.}(2022){Li}, {Deheuvels}, {Ballot}, \& {Ligni{\`e}res}}]{LiGang22}
{Li}, G., {Deheuvels}, S., {Ballot}, J., \& {Ligni{\`e}res}, F. 2022, \nat, 610, 43

\bibitem[{{Li} {et~al.}(2023){Li}, {Deheuvels}, {Li}, {Ballot}, \& {Ligni{\`e}res}}]{LiGang23}
{Li}, G., {Deheuvels}, S., {Li}, T., {Ballot}, J., \& {Ligni{\`e}res}, F. 2023, \aap, 680, A26

\bibitem[{{Lindsay} {et~al.}(2022){Lindsay}, {Ong}, \& {Basu}}]{Lindsay22}
{Lindsay}, C.~J., {Ong}, J.~M.~J., \& {Basu}, S. 2022, \apj, 931, 116

\bibitem[{Manning {et~al.}(2008)Manning, Raghavan, \& Sch{\"u}tze}]{Manning08}
Manning, C.~D., Raghavan, P., \& Sch{\"u}tze, H. 2008, Introduction to Information Retrieval (Cambridge, UK: Cambridge University Press)

\bibitem[{{Marques} {et~al.}(2013){Marques}, {Goupil}, {Lebreton}, {Talon}, {Palacios}, {Belkacem}, {Ouazzani}, {Mosser}, {Moya}, {Morel}, {Pichon}, {Mathis}, {Zahn}, {Turck-Chi{\`e}ze}, \& {Nghiem}}]{Marques13}
{Marques}, J.~P., {Goupil}, M.~J., {Lebreton}, Y., {et~al.} 2013, \aap, 549, A74

\bibitem[{{Mathis} \& {Bugnet}(2023)}]{Mathis23}
{Mathis}, S. \& {Bugnet}, L. 2023, \aap, 676, L9

\bibitem[{{Matteuzzi} {et~al.}(2025){Matteuzzi}, {Buldgen}, {Dupret}, {Miglio}, {Panier}, \& {van Rossem}}]{Matteuzzi25}
{Matteuzzi}, M., {Buldgen}, G., {Dupret}, M.-A., {et~al.} 2025, \aap, 700, A261

\bibitem[{{Miglio} {et~al.}(2010){Miglio}, {Montalb{\'a}n}, {Carrier}, {De Ridder}, {Mosser}, {Eggenberger}, {Scuflaire}, {Ventura}, {D'Antona}, {Noels}, \& {Baglin}}]{Miglio10}
{Miglio}, A., {Montalb{\'a}n}, J., {Carrier}, F., {et~al.} 2010, \aap, 520, L6

\bibitem[{{Mosser} \& {Appourchaux}(2009)}]{Mosser09}
{Mosser}, B. \& {Appourchaux}, T. 2009, \aap, 508, 877

\bibitem[{{Mosser} {et~al.}(2011a){Mosser}, {Belkacem}, {Goupil}, {Michel}, {Elsworth}, {Barban}, {Kallinger}, {Hekker}, {De Ridder}, {Samadi}, {Baudin}, {Pinheiro}, {Auvergne}, {Baglin}, \& {Catala}}]{Mosser11UP}
{Mosser}, B., {Belkacem}, K., {Goupil}, M.~J., {et~al.} 2011a, \aap, 525, L9

\bibitem[{{Mosser} {et~al.}(2017){Mosser}, {Belkacem}, {Pin{\c{c}}on}, {Takata}, {Vrard}, {Barban}, {Goupil}, {Kallinger}, \& {Samadi}}]{Mosser17}
{Mosser}, B., {Belkacem}, K., {Pin{\c{c}}on}, C., {et~al.} 2017, \aap, 598, A62

\bibitem[{{Mosser} {et~al.}(2014){Mosser}, {Benomar}, {Belkacem}, {Goupil}, {Lagarde}, {Michel}, {Lebreton}, {Stello}, {Vrard}, {Barban}, {Bedding}, {Deheuvels}, {Chaplin}, {De Ridder}, {Elsworth}, {Montalban}, {Noels}, {Ouazzani}, {Samadi}, {White}, \& {Kjeldsen}}]{Mosser14}
{Mosser}, B., {Benomar}, O., {Belkacem}, K., {et~al.} 2014, \aap, 572, L5

\bibitem[{{Mosser} {et~al.}(2024){Mosser}, {Dr{\'e}au}, {Pin{\c{c}}on}, {Deheuvels}, {Belkacem}, {Lebreton}, {Goupil}, \& {Michel}}]{Mosser24}
{Mosser}, B., {Dr{\'e}au}, G., {Pin{\c{c}}on}, C., {et~al.} 2024, \aap, 681, L20

\bibitem[{{Mosser} {et~al.}(2012{\natexlab{a}}){Mosser}, {Elsworth}, {Hekker}, {Huber}, {Kallinger}, {Mathur}, {Belkacem}, {Goupil}, {Samadi}, {Barban}, {Bedding}, {Chaplin}, {Garc{\'\i}a}, {Stello}, {De Ridder}, {Middour}, {Morris}, \& {Quintana}}]{Mosser12}
{Mosser}, B., {Elsworth}, Y., {Hekker}, S., {et~al.} 2012{\natexlab{a}}, \aap, 537, A30

\bibitem[{{Mosser} {et~al.}(2018){Mosser}, {Gehan}, {Belkacem}, {Samadi}, {Michel}, \& {Goupil}}]{Mosser18}
{Mosser}, B., {Gehan}, C., {Belkacem}, K., {et~al.} 2018, \aap, 618, A109

\bibitem[{{Mosser} {et~al.}(2012{\natexlab{b}}){Mosser}, {Goupil}, {Belkacem}, {Marques}, {Beck}, {Bloemen}, {De Ridder}, {Barban}, {Deheuvels}, {Elsworth}, {Hekker}, {Kallinger}, {Ouazzani}, {Pinsonneault}, {Samadi}, {Stello}, {Garc{\'\i}a}, {Klaus}, {Li}, {Mathur}, \& {Morris}}]{Mosser12b}
{Mosser}, B., {Goupil}, M.~J., {Belkacem}, K., {et~al.} 2012{\natexlab{b}}, \aap, 548, A10

\bibitem[{{Mosser} {et~al.}(2012{\natexlab{c}}){Mosser}, {Goupil}, {Belkacem}, {Michel}, {Stello}, {Marques}, {Elsworth}, {Barban}, {Beck}, {Bedding}, {De Ridder}, {Garc{\'\i}a}, {Hekker}, {Kallinger}, {Samadi}, {Stumpe}, {Barclay}, \& {Burke}}]{Mosser12c}
{Mosser}, B., {Goupil}, M.~J., {Belkacem}, K., {et~al.} 2012{\natexlab{c}}, \aap, 540, A143

\bibitem[{{Mosser} {et~al.}(2015){Mosser}, {Vrard}, {Belkacem}, {Deheuvels}, \& {Goupil}}]{Mosser15}
{Mosser}, B., {Vrard}, M., {Belkacem}, K., {Deheuvels}, S., \& {Goupil}, M.~J. 2015, \aap, 584, A50

\bibitem[{{Moyano} {et~al.}(2022){Moyano}, {Eggenberger}, {Meynet}, {Gehan}, {Mosser}, {Buldgen}, \& {Salmon}}]{Moyano22}
{Moyano}, F.~D., {Eggenberger}, P., {Meynet}, G., {et~al.} 2022, \aap, 663, A180

\bibitem[{{Nielsen} {et~al.}(2021){Nielsen}, {Davies}, {Ball}, {Lyttle}, {Li}, {Hall}, {Chaplin}, {Gaulme}, {Carboneau}, {Ong}, {Garc{\'\i}a}, {Mosser}, {Roxburgh}, {Corsaro}, {Benomar}, {Moya}, \& {Lund}}]{Nielsen21}
{Nielsen}, M.~B., {Davies}, G.~R., {Ball}, W.~H., {et~al.} 2021, \aj, 161, 62

\bibitem[{{Nielsen} {et~al.}(2025){Nielsen}, {Ong}, {Hatt}, {Davies}, {Chaplin}, {Hookway}, {Stokholm}, {Scutt}, {Lund}, \& {Garc{\'\i}a}}]{Nielsen25}
{Nielsen}, M.~B., {Ong}, J.~M.~J., {Hatt}, E.~J., {et~al.} 2025, \aj, 169, 322

\bibitem[{{Ong} \& {Gehan}(2023)}]{Ong23}
{Ong}, J.~M.~J. \& {Gehan}, C. 2023, \apj, 946, 92

\bibitem[{{Ouazzani} {et~al.}(2013){Ouazzani}, {Goupil}, {Dupret}, \& {Marques}}]{Ouazzani13}
{Ouazzani}, R.-M., {Goupil}, M.~J., {Dupret}, M.-A., \& {Marques}, J.~P. 2013, \aap, 554, A80

\bibitem[{{Pinsonneault} {et~al.}(2018){Pinsonneault}, {Elsworth}, {Tayar}, {Serenelli}, {Stello}, {Zinn}, {Mathur}, {Garc{\'\i}a}, {Johnson}, {Hekker}, {Huber}, {Kallinger}, {M{\'e}sz{\'a}ros}, {Mosser}, {Stassun}, {Girardi}, {Rodrigues}, {Silva Aguirre}, {An}, {Basu}, {Chaplin}, {Corsaro}, {Cunha}, {Garc{\'\i}a-Hern{\'a}ndez}, {Holtzman}, {J{\"o}nsson}, {Shetrone}, {Smith}, {Sobeck}, {Stringfellow}, {Zamora}, {Beers}, {Fern{\'a}ndez-Trincado}, {Frinchaboy}, {Hearty}, \& {Nitschelm}}]{Pinsonneault18}
{Pinsonneault}, M.~H., {Elsworth}, Y.~P., {Tayar}, J., {et~al.} 2018, \apjs, 239, 32

\bibitem[{{Pinsonneault} {et~al.}(2025){Pinsonneault}, {Zinn}, {Tayar}, {Serenelli}, {Garc{\'\i}a}, {Mathur}, {Vrard}, {Elsworth}, {Mosser}, {Stello}, {Bell}, {Bugnet}, {Corsaro}, {Gaulme}, {Hekker}, {Hon}, {Huber}, {Kallinger}, {Cao}, {Johnson}, {Liagre}, {Patton}, {Santos}, {Basu}, {Beck}, {Beers}, {Chaplin}, {Cunha}, {Frinchaboy}, {Girardi}, {Godoy-Rivera}, {Holtzman}, {J{\"o}nsson}, {M{\'e}sz{\'a}ros}, {Reyes}, {Rix}, {Shetrone}, {Smith}, {Spoo}, {Stassun}, \& {Wang}}]{APOKASC3}
{Pinsonneault}, M.~H., {Zinn}, J.~C., {Tayar}, J., {et~al.} 2025, \apjs, 276, 69

\bibitem[{{Rauer} {et~al.}(2025){Rauer}, {Aerts}, {Cabrera}, {Deleuil}, {Erikson}, {Gizon}, {Goupil}, {Heras}, {Walloschek}, {Lorenzo-Alvarez}, {Marliani}, {Martin-Garcia}, {Mas-Hesse}, {O'Rourke}, {Osborn}, {Pagano}, {Piotto}, {Pollacco}, {Ragazzoni}, {Ramsay}, {Udry}, {Appourchaux}, {Benz}, {Brandeker}, {G{\"u}del}, {Janot-Pacheco}, {Kabath}, {Kjeldsen}, {Min}, {Santos}, {Smith}, {Suarez}, {Werner}, {Aboudan}, {Abreu}, {Acu{\~n}a}, {Adams}, {Adibekyan}, {Affer}, {Agneray}, {Agnor}, {Aguirre B{\o}rsen-Koch}, {Ahmed}, {Aigrain}, {Al-Bahlawan}, {Alcacera Gil}, {Alei}, {Alencar}, {Alexander}, {Alfonso-Garz{\'o}n}, {Alibert}, {Allende Prieto}, {Almeida}, {Alonso Sobrino}, {Altavilla}, {Althaus}, {Alvarez Trujillo}, {Amarsi}, {Ammler-von Eiff}, {Am{\^o}res}, {Andrade}, {Antoniadis-Karnavas}, {Ant{\'o}nio}, {Aparicio del Moral}, {Appolloni}, {Arena}, {Armstrong}, {Aroca Aliaga}, {Asplund}, {Audenaert}, {Auricchio}, {Avelino}, {Baeke}, {Bailli{\'e}}, {Balado}, {Ballber Balaguer{\'o}}, {Balestra}, {Ball}, {Ballans}, {Ballot}, {Barban}, {Barbary}, {Barbieri}, {Barcel{\'o} Forteza}, {Barker}, {Barklem}, {Barnes}, {Barrado Navascues}, {Barragan}, {Baruteau}, {Basu}, {Baudin}, {Baumeister}, {Bayliss}, {Bazot}, {Beck}, {Belkacem}, {Bellinger}, {Benatti}, {Benomar}, {B{\'e}rard}, {Bergemann}, {Bergomi}, {Bernardo}, {Biazzo}, {Bignamini}, {Bigot}, {Billot}, {Binet}, {Biondi}, {Biondi}, {Birch}, {Bitsch}, {Bluhm Ceballos}, {B{\'o}di}, {Bogn{\'a}r}, {Boisse}, {Bolmont}, {Bonanno}, {Bonavita}, {Bonfanti}, {Bonfils}, {Bonito}, {Bonomo}, {B{\"o}rner}, {Boro Saikia}, {Borreguero Mart{\'\i}n}, {Borsa}, {Borsato}, {Bossini}, {Bouchy}, {Bou{\'e}}, {Boufleur}, {Boumier}, {Bourrier}, {Bowman}, {Bozzo}, {Bradley}, {Bray}, {Bressan}, {Breton}, {Brienza}, {Brito}, {Brogi}, {Brown}, {Brown}, {Brun}, {Bruno}, {Bruns}, {Buchhave}, {Bugnet}, {Buldgen}, {Burgess}, {Busatta}, {Busso}, {Buzasi}, {Caballero}, {Cabral}, {Cabrero Gomez}, {Calderone}, {Cameron}, {Cameron}, {Campante}, {Campos Gestal}, {Canto Martins}, {Cara}, {Carone}, {Carrasco}, {Casagrande}, {Casewell}, {Cassisi}, {Castellani}, {Castro}, {Catala}, {Catal{\'a}n Fern{\'a}ndez}, {Catelan}, {Cegla}, {Cerruti}, {Cessa}, {Chadid}, {Chaplin}, {Charpinet}, {Chiappini}, {Chiarucci}, {Chiavassa}, {Chinellato}, {Chirulli}, {Christensen-Dalsgaard}, {Church}, {Claret}, {Clarke}, {Claudi}, {Clermont}, {Coelho}, {Coelho}, {Cogato}, {Colom{\'e}}, {Condamin}, {Conde Garc{\'\i}a}, \& {Conseil}}]{Rauer25}
{Rauer}, H., {Aerts}, C., {Cabrera}, J., {et~al.} 2025, Experimental Astronomy, 59, 26

\bibitem[{{Ricker} {et~al.}(2014){Ricker}, {Winn}, {Vanderspek}, {Latham}, {Bakos}, {Bean}, {Berta-Thompson}, {Brown}, {Buchhave}, {Butler}, {Butler}, {Chaplin}, {Charbonneau}, {Christensen-Dalsgaard}, {Clampin}, {Deming}, {Doty}, {De Lee}, {Dressing}, {Dunham}, {Endl}, {Fressin}, {Ge}, {Henning}, {Holman}, {Howard}, {Ida}, {Jenkins}, {Jernigan}, {Johnson}, {Kaltenegger}, {Kawai}, {Kjeldsen}, {Laughlin}, {Levine}, {Lin}, {Lissauer}, {MacQueen}, {Marcy}, {McCullough}, {Morton}, {Narita}, {Paegert}, {Palle}, {Pepe}, {Pepper}, {Quirrenbach}, {Rinehart}, {Sasselov}, {Sato}, {Seager}, {Sozzetti}, {Stassun}, {Sullivan}, {Szentgyorgyi}, {Torres}, {Udry}, \& {Villasenor}}]{Ricker14}
{Ricker}, G.~R., {Winn}, J.~N., {Vanderspek}, R., {et~al.} 2014, in Society of Photo-Optical Instrumentation Engineers (SPIE) Conference Series, Vol. 9143, Space Telescopes and Instrumentation 2014: Optical, Infrared, and Millimeter Wave, ed. J.~{Oschmann}, Jacobus~M., M.~{Clampin}, G.~G. {Fazio}, \& H.~A. {MacEwen}, 914320

\bibitem[{{Schimak} {et~al.}(2026){Schimak}, {Bedding}, {Crawford}, {Beck}, {Li}, {Huber}, {Ong}, {Montet}, {Pedersen}, {Grossmann}, {Mathur}, \& {Garc{\'\i}a}}]{Schimak26}
{Schimak}, L.~S., {Bedding}, T.~R., {Crawford}, C.~L., {et~al.} 2026, \mnras, 546, stag151

\bibitem[{Schreiber {et~al.}(2003)Schreiber, Fellous, Whitmer, Tiesinga, \& Sejnowski}]{Schreiber03}
Schreiber, S., Fellous, J., Whitmer, D., Tiesinga, P., \& Sejnowski, T. 2003, Neurocomputing, 52-54, 925, computational Neuroscience: Trends in Research 2003

\bibitem[{{Shibahashi}(1979)}]{Shibahashi79}
{Shibahashi}, H. 1979, \pasj, 31, 87

\bibitem[{{Silva Aguirre} {et~al.}(2014){Silva Aguirre}, {Ruchti}, {Hekker}, {Cassisi}, {Christensen-Dalsgaard}, {Datta}, {Jendreieck}, {Jessen-Hansen}, {Mazumdar}, {Mosser}, {Stello}, {Beck}, \& {de Ridder}}]{SilvaAguirre14}
{Silva Aguirre}, V., {Ruchti}, G.~R., {Hekker}, S., {et~al.} 2014, \apjl, 784, L16

\bibitem[{{Tassoul}(1980)}]{Tassoul1980}
{Tassoul}, M. 1980, \apjs, 43, 469

\bibitem[{{Tayar} {et~al.}(2019){Tayar}, {Beck}, {Pinsonneault}, {Garc{\'\i}a}, \& {Mathur}}]{Tayar19}
{Tayar}, J., {Beck}, P.~G., {Pinsonneault}, M.~H., {Garc{\'\i}a}, R.~A., \& {Mathur}, S. 2019, \apj, 887, 203

\bibitem[{{Theme{\ss}l} {et~al.}(2018a){Theme{\ss}l}, {Hekker}, {Southworth}, {Beck}, {Pavlovski}, {Tkachenko}, {Angelou}, {Ball}, {Barban}, {Corsaro}, {Elsworth}, {Handberg}, \& {Kallinger}}]{Themessl18a}
{Theme{\ss}l}, N., {Hekker}, S., {Southworth}, J., {et~al.} 2018a, \mnras, 478, 4669

\bibitem[{{Triana} {et~al.}(2017){Triana}, {Corsaro}, {De Ridder}, {Bonanno}, {P{\'e}rez Hern{\'a}ndez}, \& {Garc{\'\i}a}}]{Triana17}
{Triana}, S.~A., {Corsaro}, E., {De Ridder}, J., {et~al.} 2017, \aap, 602, A62

\bibitem[{{Unno} {et~al.}(1989){Unno}, {Osaki}, {Ando}, {Saio}, \& {Shibahashi}}]{Unno89}
{Unno}, W., {Osaki}, Y., {Ando}, H., {Saio}, H., \& {Shibahashi}, H. 1989, {Nonradial oscillations of stars}

\bibitem[{van Rossum(2001)}]{vanRossum01}
van Rossum, M.~C. 2001, Neural Comput., 13, 751

\bibitem[{Victor \& Purpura(1996)}]{VictorPurpura96}
Victor, J.~D. \& Purpura, K.~P. 1996, J. Neurophysiology, 76, 1310

\bibitem[{Victor \& Purpura(1997)}]{VictorPurpura97}
Victor, J.~D. \& Purpura, K.~P. 1997, Network: Computation in Neural Systems, 8, 127

\bibitem[{{Vrard} {et~al.}(2022){Vrard}, {Cunha}, {Bossini}, {Avelino}, {Corsaro}, \& {Mosser}}]{Vrard22}
{Vrard}, M., {Cunha}, M.~S., {Bossini}, D., {et~al.} 2022, Nature Communications, 13, 7553

\bibitem[{{Vrard} {et~al.}(2016){Vrard}, {Mosser}, \& {Samadi}}]{Vrard16}
{Vrard}, M., {Mosser}, B., \& {Samadi}, R. 2016, \aap, 588, A87

\bibitem[{{Vrard} {et~al.}(2025){Vrard}, {Pinsonneault}, {Elsworth}, {Hon}, {Kallinger}, {Kuszlewicz}, {Mosser}, {Garc{\'\i}a}, {Tayar}, {Bennett}, {Cao}, {Hekker}, {Loyer}, {Mathur}, \& {Stello}}]{Vrard25}
{Vrard}, M., {Pinsonneault}, M.~H., {Elsworth}, Y., {et~al.} 2025, \aap, 697, A165

\bibitem[{{Wang} {et~al.}(2025){Wang}, {Lin}, {Li}, {Wu}, \& {Li}}]{Wang25}
{Wang}, Y., {Lin}, G., {Li}, Y., {Wu}, T., \& {Li}, Y. 2025, \apj, 985, 8

\bibitem[{{Yu} {et~al.}(2018){Yu}, {Huber}, {Bedding}, {Stello}, {Hon}, {Murphy}, \& {Khanna}}]{Yu18}
{Yu}, J., {Huber}, D., {Bedding}, T.~R., {et~al.} 2018, \apjs, 236, 42

\bibitem[{{Zhang} {et~al.}(2022){Zhang}, {Cai}, {Li}, \& {Wu}}]{Zhang22}
{Zhang}, X., {Cai}, T., {Li}, Y., \& {Wu}, T. 2022, \apj, 931, 64

\bibitem[{{Zhang} {et~al.}(2018){Zhang}, {Wu}, \& {Li}}]{Zhang18}
{Zhang}, X., {Wu}, T., \& {Li}, Y. 2018, \apj, 855, 16

\end{thebibliography}

\begin{appendix}
   \section{Scaling relations and initial parameter estimates}\label{appendix:scaling_relations}
      To generate realistic synthetic spectra, we used empirical scaling relations and analytical fits to initialise the parameters of the mixed-mode equation. These initial estimates are parameterised as a function of the frequency of maximum power, $\nu_{\rm max}$, or the large frequency separation, $\Delta\nu$. Table~\ref{tab:scaling_relations} summarises the relations; those marked as ``CAPASS'' were derived via linear fits to the results in the CAPASS catalogue (Espinoza-Rojas et al., in prep.), adopting evolutionary states from \citet{Vrard25}.

      \begin{table}[h]
         \centering
         \caption{Scaling relations used to initialise the asymptotic parameters.}\label{tab:scaling_relations}
         \renewcommand{\arraystretch}{1.3}
         \begin{tabular}{ccl}
            \hline\hline
            Parameter & Evol. state & Relation \\
            \hline
            
            $\Delta\nu$ & RGB & $0.260 \, \nu_{\max}^{0.761}$ $^{a}$ \\
            & CHeB & $0.0567 \, \nu_{\max} + 2.135$ $^{b}$ \\
            $\epsilon_p$ & RGB & $0.546 \log_{10}(\Delta\nu) + 0.634$ $^{c}$ \\
            & CHeB & $0.0250 \, \Delta\nu + 0.917$ $^{b}$ \\
            $\alpha$ & Both & $10^{-(0.0526 \, \Delta\nu + 1.54)}$ $^{d}$ \\
            
            $\Delta\Pi_1$ & RGB & $2.176 \, \Delta\nu + 52.84$ $^{f}$ \\
            & CHeB & $-2.752 \, \Delta\nu + 277.50$ $^{b}$ \\
            $q$ & RGB & $0.0024 \, \Delta\nu + 0.104$ $^{g}$ \\
            & CHeB & $0.3$ $^{h}$ \\
            $\epsilon_g$ & Both$^*$ & $0.00540 \, \Delta\nu - 0.3132$ $^{i}$ \\
            $d_{01}$ & Both$^*$ & $0.00263 \, \Delta\nu - 0.04259$ $^{i}$ \\
            $\delta\nu_{\rm rot}$ & RGB  & $352.4$~nHz \\
                            &CHeB & $51.0$~nHz \\
            \hline
         \end{tabular}
         
         \tablefoot{ $\nu_{\max}$ and $\Delta\nu$ are in $\mu$Hz, and $\Delta\Pi_1$ is in seconds. $q^{\rm CHeB}$, $\delta\nu_{\rm rot}^{\rm RGB}$, and $\delta\nu_{\rm rot}^{\rm CHeB}$ represent the median of the values reported in \citet{Mosser17}, \citet{Gehan18}, and \citet{Mosser24}, respectively. \small $^*$Applied to CHeB stars as well, as no specific estimates have yet been reported for this evolutionary state. \small $^{a}$\citet{Hekker11}; $^{b}$CAPASS; $^{c}$\citet{Mosser11UP}; $^{d}$\citet{Hekker26}; $^{e}$\citet{Huber10}; \small $^{f}$Linear fit to \citet{Mosser14}; $^{g}$Linear fit to \citet{Mosser17}; $^{h}$\citet{Mosser17}; \small $^{i}$Linear fit to \citet{LiGang24}.}
      \end{table}

      \subsection{Rotational splittings}

         We assigned rotational splittings, $\delta\nu_{\rm rot}$, to match specific levels of spectral complexity, which we quantified using the fraction of crossovers in the frequency spectrum (see the crossover ratio, $R$, in Sect.~\ref{subsec:grid}). We defined four rotational regimes: $R\approx 0.0$, $0.2$, $0.4$, and $0.6$. For each regime, the splitting (in nHz) is parameterised as a second-order polynomial of $\nu_{\max}$:
         \begin{equation}
         \delta\nu_{\rm rot} = a \nu_{\max}^2 + b \nu_{\max} + c.
         \end{equation}
         This relation was applied to all synthetic CHeB stars, and to synthetic RGB stars with $\nu_{\max} \geq 50~\mu$Hz. The polynomial coefficients ($a, b, c$) for each regime and evolutionary state are provided in Table~\ref{tab:rotation_coeffs}.

         \begin{table}[h]
         \centering
         \caption{Coefficients of the second-order polynomial for $\delta\nu_{\rm rot}$ as a function of $\nu_{\max}$.}\label{tab:rotation_coeffs}
         \begin{tabular}{llccc}
         \hline\hline
         Evol. state & Level & $a$~[$1/\mu$Hz] & $b$ & $c$~[$\mu$Hz] \\
         \hline
         RGB  & $R\approx 0.0$ & $3.196 \times 10^{-2}$ & $-2.354$ & $72.31$ \\
            & $R\approx 0.2$ & $3.255 \times 10^{-2}$ & $-1.418$ & $31.54$ \\
            & $R\approx 0.4$ & $3.634 \times 10^{-2}$ & $-0.994$ & $13.88$ \\
            & $R\approx 0.6$ & $3.966 \times 10^{-2}$ & $-0.198$ & $-14.83$ \\
         \hline
         CHeB & $R\approx 0.0$ & $0.079$ & $-1.541$ & $7.086$ \\
            & $R\approx 0.2$ & $0.099$ & $-1.614$ & $6.377$ \\
            & $R\approx 0.4$ & $0.127$ & $-1.757$ & $4.690$ \\
            & $R\approx 0.6$ & $0.164$ & $-1.768$ & $1.777$ \\
         \hline
         \end{tabular}
         \end{table}

   \section{Gaussian kernel widths for synthetic stars}\label{appendix:sigma}
      The Gaussian kernel widths ($\sigma_{k}$) used to construct the multi-bandwidth score for each synthetic star are listed in Table~\ref{table:sigma_list}. These bounds were empirically defined based on the distribution of the line widths (reported in CAPASS) of the dipole mode candidates for each evolutionary state. The size of $\hat{\sigma}$ can be chosen freely, taking into account that larger sets of $\sigma$ increase the computation time. We recommend using at least two values, i.e. the lower and upper bounds.

      \renewcommand{\arraystretch}{1.2}
      \begin{table}
         \caption{Empirically determined Gaussian kernel widths used in $\tilde{\mathcal{S}}_{\rm mbw}$ of synthetic stars for different $\nu_{\rm max}$ and evolutionary states.} \label{table:sigma_list}   
            \centering                     
            \begin{tabular}{ccc|ccc}
               \hline\hline
                     & \multicolumn{2}{c|}{RGB} & \multicolumn{3}{c}{CHeB} \\
               $\nu_{\rm max}$  & $\sigma_{\rm low}$ & $\sigma_{\rm up}$ & $\nu_{\rm max}$ & $\sigma_{\rm low}$ & $\sigma_{\rm up}$ \\\relax
               [$\mu$Hz]        & [$\mu$Hz]          & [$\mu$Hz]         & [$\mu$Hz]       & [$\mu$Hz]          & [$\mu$Hz]         \\
               \hline
               20               & 0.007              & 0.01              & 20              & 0.007              & 0.01              \\
               40               & 0.007              & 0.01              & 30              & 0.007              & 0.01              \\
               60               & 0.007              & 0.025             & 40              & 0.007              & 0.025             \\
               80               & 0.007              & 0.025             & 50              & 0.007              & 0.025             \\
               100              & 0.007              & 0.025             & 60              & 0.013              & 0.05              \\
               150              & 0.007              & 0.1               & 70              & 0.013              & 0.05              \\
               200              & 0.007              & 0.1               & 80              & 0.013              & 0.05              \\
               250              & 0.013              & 0.2               & 100             & 0.013              & 0.1               \\
               \hline
            \end{tabular}
      \end{table}
      \renewcommand{\arraystretch}{1.2}   
      
   \section{Description of the genetic algorithm}\label{appendix:GA}
      Genetic algorithms belong to a family of optimisation methods inspired by the biological process of natural selection. In this work, we used the GA implementation provided in the {\tt pyGAD} Python package \citep{pygad}. 
      
      The central idea is to create a random ``population'' of candidate solutions and evolve them over many generations until they closely match the observed frequency spectrum. In this work, an individual ``candidate'' (or ``chromosome'') is a specific combination of the parameters of the asymptotic equation (e.g. $\nu_{\rm max}, \Delta\nu, \epsilon_{p}, \alpha, \Delta\Pi_{1}, q, \epsilon_{g}$, and $\delta\nu_{\rm rot}$). We refer to the values of each of these parameters as ``traits''.

      The method depends entirely on the fitness function, as it acts as the ``environment'', deciding which parameter combinations are good enough to survive. We used the multi-bandwidth overlap score ($\tilde{\mathcal{S}}_{\rm mbw}$) introduced in Sect.~\ref{subsec:Smbw} as our fitness function. A higher score means the theoretical frequencies generated by those parameters closely align with the observed data.

      \subsection{Process}
         The algorithm starts with an initial population of 200 candidates. To evolve this population into the next generation, {\tt pyGAD} performs the following sequential operations:

         \begin{itemize}
            \item Elitism: Before making any changes, the top 5\% of the best-performing candidates are copied directly into the next generation. This guarantees that the algorithm never loses its best solution.
            
            \item Parent selection: To create the rest of the new generation, the algorithm must choose ``parents''. We used a ``tournament selection'' method with a size of $K=4$. The algorithm randomly picks four candidates from the initial population and selects the one with the highest overlap score ($\tilde{\mathcal{S}}_{\rm mbw}$). This ensures that good candidates reproduce frequently, but it still gives slightly worse candidates a chance to be selected, which prevents the algorithm from getting stuck in local maxima. We chose 25\% of the initial population as parents.
            
            \item Crossover: Selected parents are paired up to produce ``offspring''. They undergo a ``two-point crossover'' with a 90\% probability. To understand this process, imagine the parameter list of each parent laid out in a row; the algorithm cuts the lists at two random points and swaps the middle sections between the parents. This effectively mixes the traits of different chromosomes to generate offspring that make up 95\% of the next generation.
            
            \item Mutation: To introduce new traits and explore unseen regions of the parameter space, the offspring undergo random mutations while the elite sample remains unperturbed. For this, we used ``adaptive'' mutation, which scales based on the quality of the individual offspring relative to the current population. If an offspring's fitness is below the generation's average, the algorithm heavily mutates seven of its parameters to allow for broad exploration and escape local maxima. However, for above-average offspring, the mutation rate automatically decreases to just two parameters. This dynamic approach allows the algorithm to aggressively improve poor candidates while carefully fine-tuning the best results without completely deteriorating them.
            
         \end{itemize}
      
      \renewcommand{\arraystretch}{1.2}
         \begin{table}[h!]
         \centering
         \caption{Hyperparameters used for the genetic algorithm.}
         \begin{tabular}{lc}
         \hline\hline
         Parameter & Value \\
         \hline
         Maximum generations & 3000 \\
         Population size & 200 \\
         Selection method & Tournament \\
         Tournament size ($K$) & 4 \\
         Elitism (parents kept) & 5\% (10 individuals) \\
         Crossover type & Two-point \\
         Crossover probability & 0.9 (90\%) \\
         Mutation type & Adaptive \\
         Mutation bounds (max, min genes) & 7, 2 \\
         \hline
         \end{tabular}\label{table:ga_parameters}
         \end{table}
      \renewcommand{\arraystretch}{1.0}

      \subsection{Stopping criteria}
      We set the GA to run for a maximum of 3000 generations, which we empirically found to be sufficient to ensure a broad exploration of the parameter space in this context. However, to save computational time, the code is designed to stop early under two specific conditions:
      \begin{enumerate}
         \item Target reached: The algorithm immediately stops if it finds a fit that achieves a perfect overlap score ($\tilde{\mathcal{S}}_{\rm mbw} = 1.0$).
         \item Saturation: If the score of the best fit does not improve for 1000 consecutive generations, the algorithm assumes it has converged on the final solution and stops the evolution process.
      \end{enumerate}

      A complete summary of the hyperparameters used to configure the GA in this work is provided in Table~\ref{table:ga_parameters}.

   \section{Figures for the statistical validation of synthetic stars with multiplets}\label{appendix:validation_extra_figures}
\begin{figure*}[h!]
         \centering
         \includegraphics[width=0.95\textwidth]{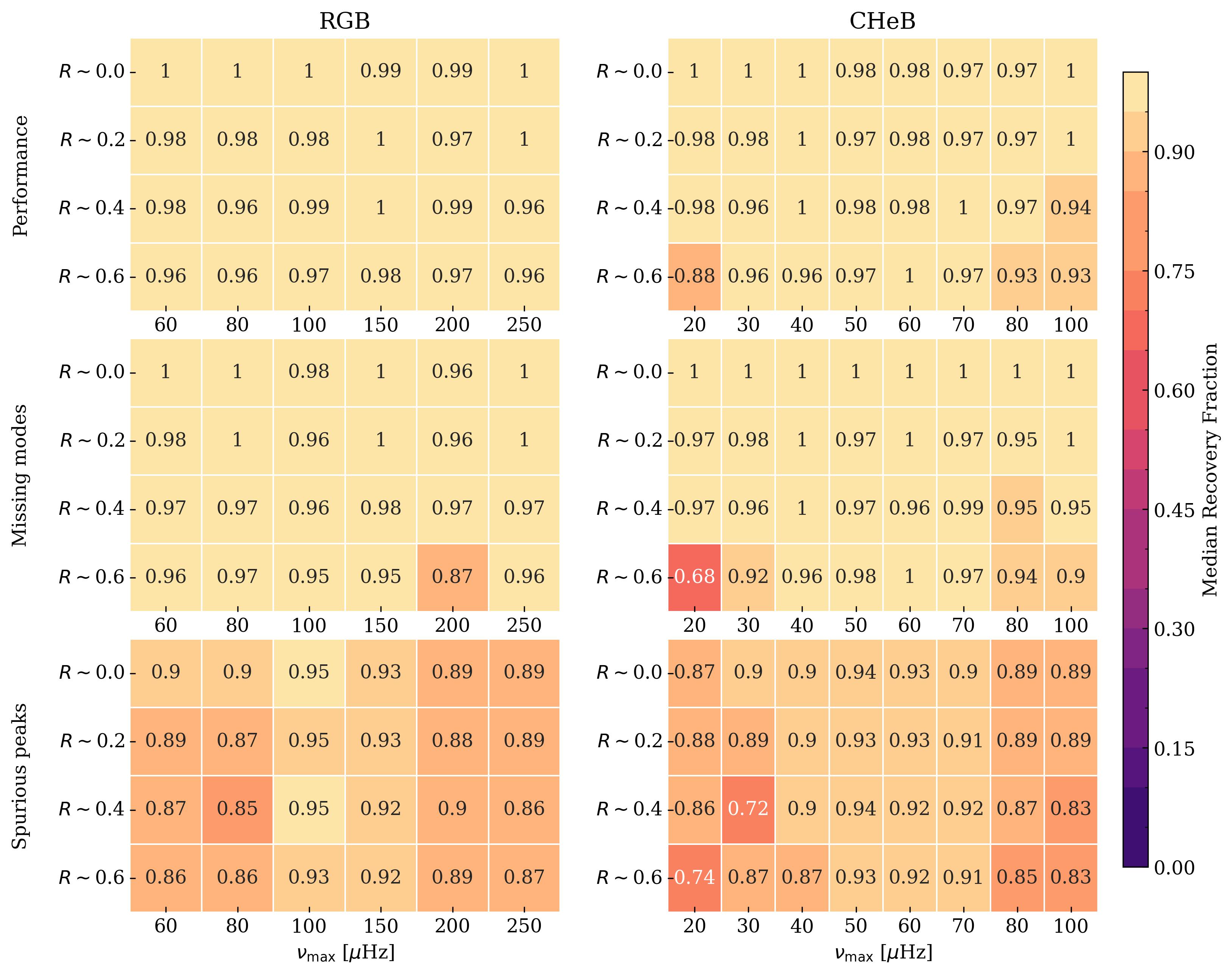}
         \caption{Validation results for synthetic stars with doublets. The grid is divided by evolutionary state (RGB on the left and CHeB on the right) and by test: performance (P, top row), missing modes (MM, middle row), and spurious peaks (SP, bottom row). Within each panel, the algorithm's performance is evaluated as a function of $\nu_{\rm max}$ (x-axis) and the crossover ratio ($R$, y-axis). Each cell is colour-coded by the median recovery fraction.} \label{fig:heatmap_doublets}
      \end{figure*}

\begin{figure*}[h!]
         \centering
         \includegraphics[width=0.95\textwidth]{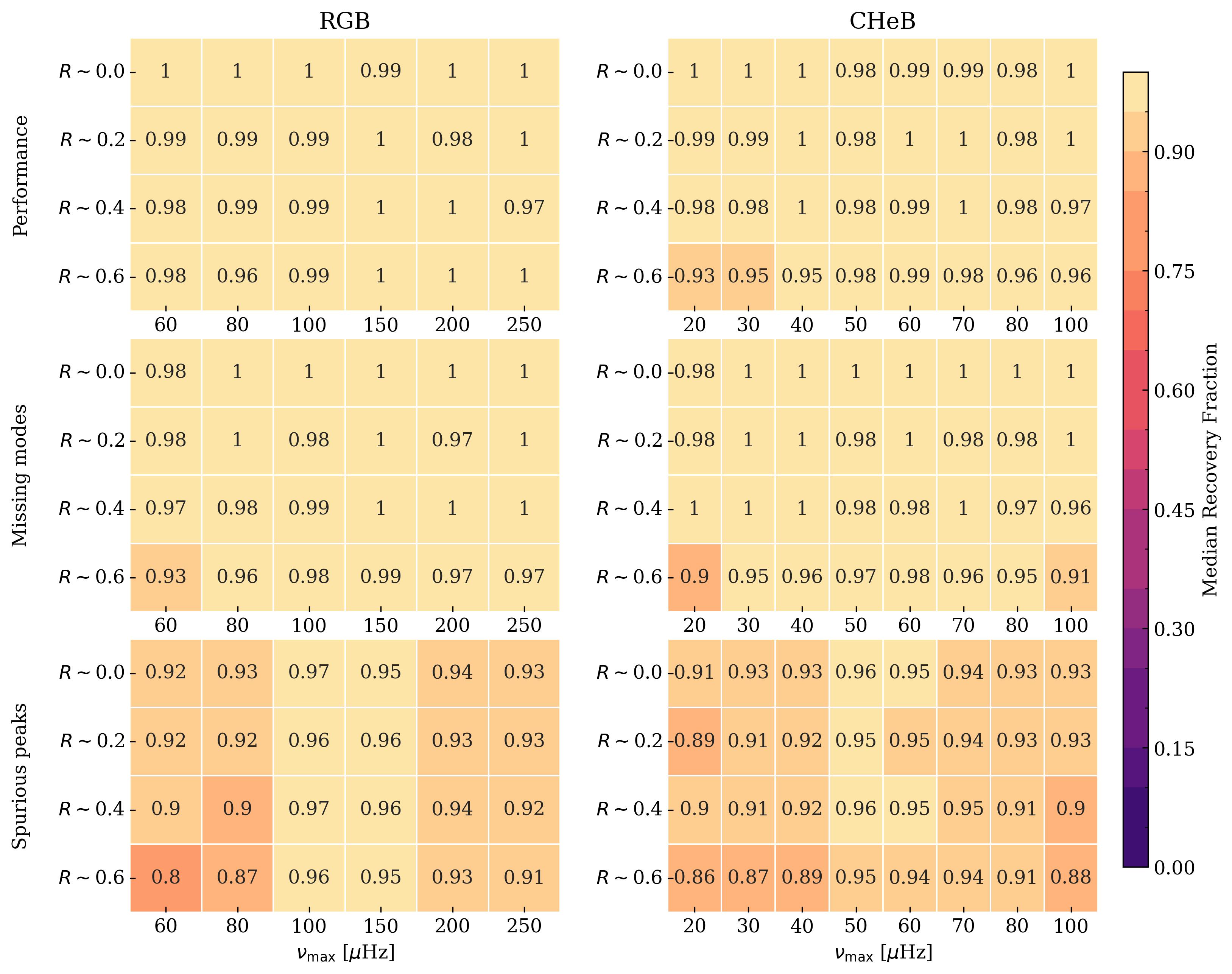}
         \caption{Same as Fig.~\ref{fig:heatmap_doublets} but for synthetic stars with triplets.}\label{fig:heatmap_triplets}
      \end{figure*}

      In this section, we present Figs.~\ref{fig:heatmap_doublets} and \ref{fig:heatmap_triplets}, which show the validation results of the synthetic stars with doublets and triplets in the same way as Fig.~\ref{fig:heatmap_singlets}.
   
   \section{Parameters of benchmark stars}\label{appendix:literature_params}
      Table~\ref{table:comparison_literature} lists the parameters of the mixed-mode equation reported for the four benchmark stars used in this work. These results are taken from \citet{SilvaAguirre14}, \citet{DiMauro16}, \citet{LiGang24}, and \citet{Mosser24}.

      \renewcommand{\arraystretch}{1.2}
      \begin{table*}[h!]
         \centering
         \setlength{\tabcolsep}{4pt}
         \caption{Comparison of asteroseismic parameters for the four benchmark red giants with the literature.}
         \begin{tabular}{lcccc}
         \hline\hline
                                 & KIC 9145955                        & KIC 12008916                       & KIC 4448777                        & KIC 5000307 \\
         \hline\hline
         $\nu_{\max}$            & $129.44$                           & $160.27$                           & $223.12$                           & $42.71$                           \\
                                 & $133.90$\tablefootmark{a}          & $163.41$\tablefootmark{a}          & $220.89$\tablefootmark{a}          & $42.46$\tablefootmark{d}          \\
                                 &                                    &                                    & $219.75$\tablefootmark{b}          & $41.99 \pm 0.62$\tablefootmark{e} \\
                                 &                                    &                                    & $216.49$\tablefootmark{c}          &                                   \\
         \hline 
         $\Delta\nu$             & $11.06$                            & $12.84$                            & $17.02$                            & $4.75$                            \\
                                 & $11.08\pm0.01$\tablefootmark{a}    & $12.92 \pm 0.01$\tablefootmark{a}  & $17.0\pm 0.005$\tablefootmark{a}   & $4.77$\tablefootmark{d}           \\
                                 &                                    &                                    & $16.96$\tablefootmark{b}           & $4.77 \pm 0.03$\tablefootmark{e}  \\
                                 &                                    &                                    & $17.02$\tablefootmark{c}           &                                   \\
         \hline
         $\alpha$                & $0.0107$                           & $0.0106$                           & $0.0073$                           & $0.0175$                          \\
                                 & $0.0081\pm0.0005$\tablefootmark{a} & $0.0081 \pm 0.0006$\tablefootmark{a}& $0.0046 \pm 0.0003$\tablefootmark{a}& --                               \\
         \hline
         $\epsilon_p$            & $1.20$                             & $1.29$                             & $0.34$                             & $1.07$                            \\
                                 & $1.14 \pm 0.01$\tablefootmark{a}   & $0.22 \pm 0.01$\tablefootmark{a}   & $0.340 \pm 0.004$\tablefootmark{a} & --                               \\
         \hline
         $d_{01}$                & $0.0156$                           & $-0.0210$                          & $0.0240$                           & $-0.0434$                         \\
                                 & $0.0059\pm0.0005$\tablefootmark{a} & $-0.136\pm0.0212$\tablefootmark{a}& $-0.0051 \pm 0.0005$\tablefootmark{a}& --                               \\
         \hline
         $\Delta\Pi_1$           & $76.74$                            & $80.04$                            & $89.36$                            & $321.37$                          \\
                                 & $76.83\pm0.02$\tablefootmark{a}   & $79.93 \pm 0.01$\tablefootmark{a}  & $89.44 \pm 0.01$\tablefootmark{a}  & $319.95$\tablefootmark{d}         \\
                                 &                                    &                                    & $89.87$\tablefootmark{b}           & $319.8 \pm 1.4$\tablefootmark{e}  \\
                                 &                                    &                                    & $89.20$\tablefootmark{c}           &                                   \\
         \hline
         $\epsilon_g$            & $0.36$                             & $0.38$                             & $0.28$                             & $0.78$                            \\
                                 & $0.24\pm 0.02$\tablefootmark{a}    & $0.49 \pm 0.01$\tablefootmark{a}   & $0.240 \pm 0.01$\tablefootmark{a} & --                               \\
         \hline
         $q$                     & $0.15$                             & $0.13$                             & $0.16$                             & $0.28$                            \\
                                 & $0.14\pm 0.003$\tablefootmark{a}   & $0.137 \pm 0.002$\tablefootmark{a} & $0.16\pm 0.001$\tablefootmark{a}& $0.25 \pm 0.03$\tablefootmark{e} \\
         \hline
         $\delta\nu_{\rm rot}$   & --                                 & $465.19$                           & $368.67$                           & --                               \\
                                 & --                                 & $429.65\pm2.61$\tablefootmark{a}  & $370.0 \pm 6.9$\tablefootmark{b}   & $50.0$\tablefootmark{d}           \\
                                 &                                    &                                    &                                    & $42 \pm 10$\tablefootmark{e}      \\
         \hline
      \end{tabular}\label{table:comparison_literature}
      \tablefoot{Frequencies ($\nu_{\max}$, $\Delta\nu$, $d_{01}$) are expressed in $\mu$Hz, the period spacing ($\Delta\Pi_1$) in seconds, and the rotational splitting ($\delta\nu_{\rm rot}$) in nHz. All other parameters are dimensionless. Values marked with superscript $(a)$ are reported by \citet{LiGang24}, $(b)$ by \citet{DiMauro16}, $(c)$ by \citet{Gehan18}, $(d)$ by \citet{SilvaAguirre14}, and $(e)$ by \citet{Mosser24}.}
      \end{table*}

   \section{Prospects for buoyancy glitches}\label{appendix:glitches}
      \begin{figure*}[h!]
         \centering
         \includegraphics[width=0.4\textwidth]{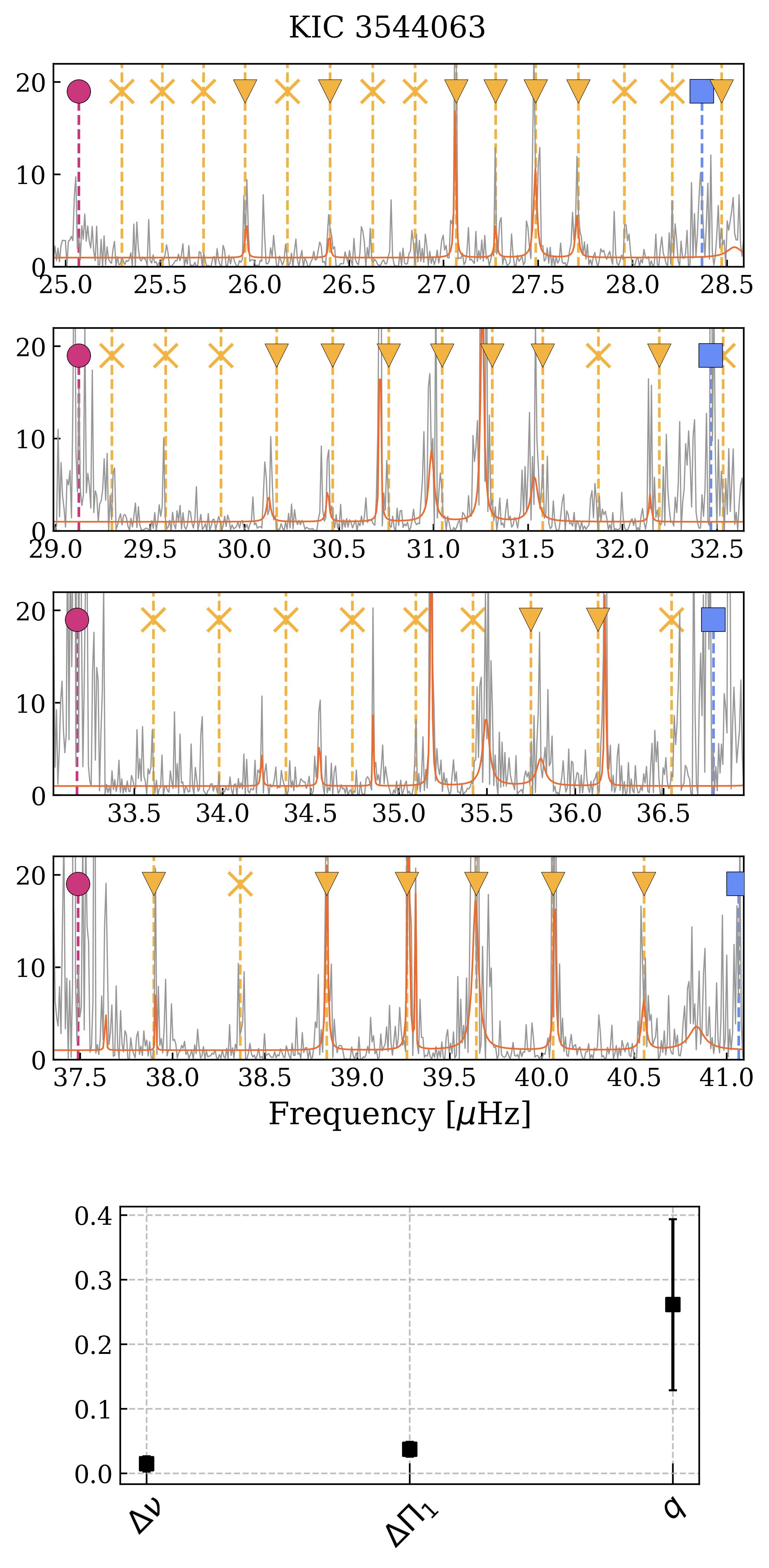}
         \includegraphics[width=0.4\textwidth]{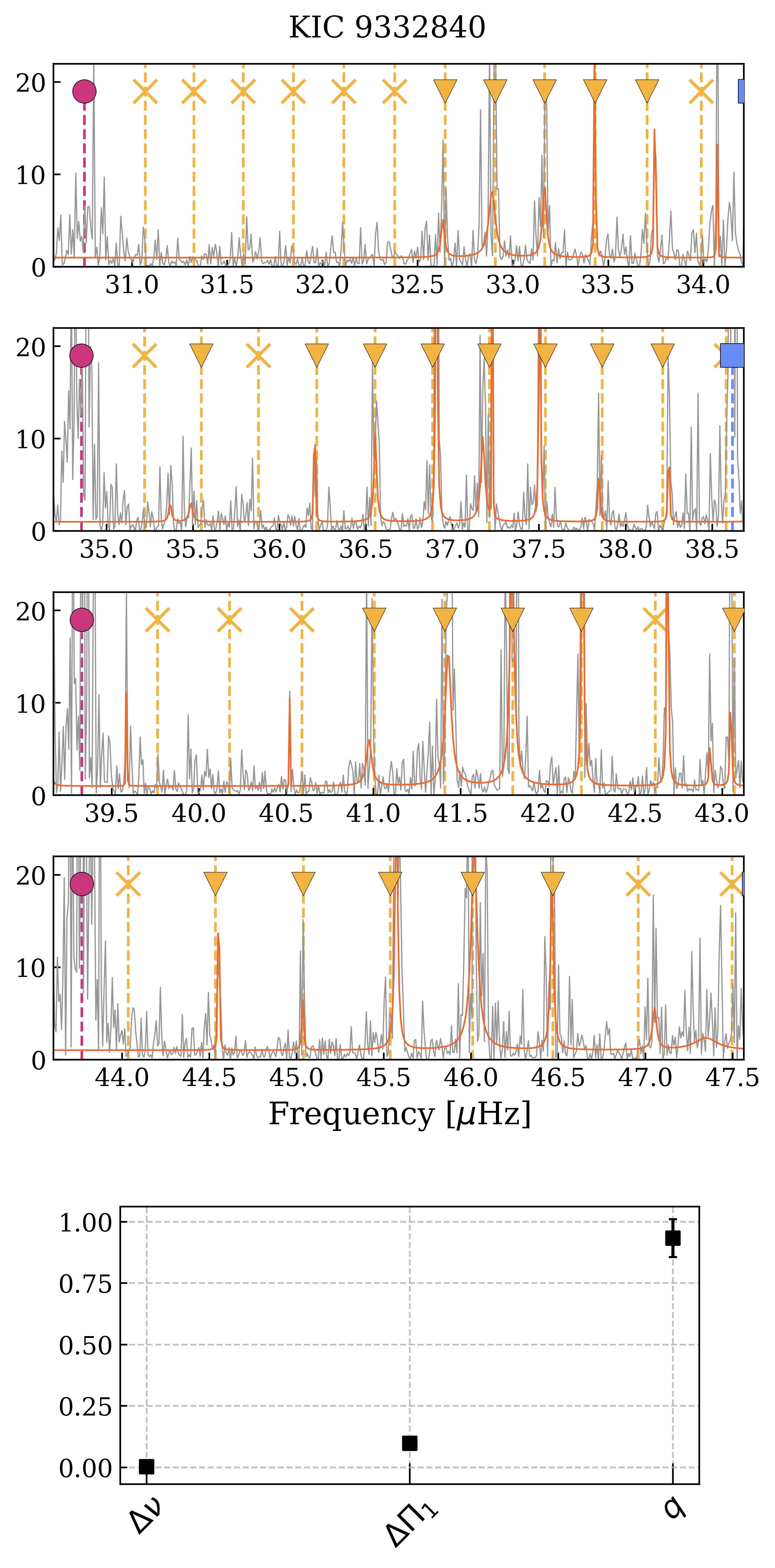}
         \caption{Same as Fig.~\ref{fig:kepler_rgs} but for KIC~3544063 and KIC~9332840.}\label{fig:glitches}
      \end{figure*}

      Oscillation frequencies can be affected by sharp structural variations in the deeper layers of stars \citep[e.g.][]{Gough90}. When these variations take place on scales smaller than the mode wavelength, they are commonly referred to as glitches. These glitches arise at the borders of convectively mixed regions, regions with chemical composition variations, or regions of ionisation \citep[e.g.][]{Miglio10,JCD14,Vrard22}. In particular, for dipole mixed modes, these glitches can be observed as a variation in the underlying period spacing of pure $g$ modes \citep[e.g.][]{Cunha15,Cunha19,Matteuzzi25,Schimak26}, deviating from the regular asymptotic pattern. Since the asymptotic dipole-mode expression used in this method does not account for structural discontinuities, we performed an initial test to check how well our approach can identify dipole modes in such cases. For this purpose, we applied our method to two clump stars, KIC~3544063 and KIC~9332840, which exhibit a core-glitch signature as reported in \citet{Vrard22}. The results are presented in Fig.~\ref{fig:glitches}.

      For KIC~9332840, our method identifies most dipole modes, missing only a few candidates. For KIC~3544063, however, it rejects many modes clearly visible in the PDS. This limitation is a consequence of applying a global fit across all radial orders. Because buoyancy glitches cause local variations in $\Delta\Pi_{1}$, the most perturbed modes can shift beyond our method's frequency tolerance and thus go undetected. This could be overcome by optimising the mixed-mode equation locally. By allowing the parameter set $\theta$ to vary per radial order, the pipeline could track these structural changes in period spacing without needing an explicit glitch model. However, a more conclusive assessment requires testing more stars, which we will present in a forthcoming paper.

\end{appendix}

\end{document}